\documentclass[aps,prd,showpacs,amssymb,nofootinbib]{revtex4}
\usepackage{graphicx}

\begin{document}
\title{Asymptotic Analysis of Field Commutators for Einstein-Rosen Gravitational Waves}

\author{J. Fernando \surname{Barbero G.}}
\email[]{jfbarbero@imaff.cfmac.csic.es} \affiliation{Instituto de
Matem\'aticas y F\'{\i}sica Fundamental, Centro de F\'{\i}sica
Miguel A. Catal\'{a}n, C.S.I.C., Serrano 113bis-121, 28006 Madrid,
Spain}
\author{Guillermo A. \surname{Mena Marug\'an}}
\email[]{mena@imaff.cfmac.csic.es} \affiliation{Instituto de
Matem\'aticas y F\'{\i}sica Fundamental, Centro de F\'{\i}sica
Miguel A. Catal\'{a}n, C.S.I.C., Serrano 113bis-121, 28006 Madrid,
Spain}
\author{Eduardo J. \surname{S. Villase\~nor}}
\email[]{eduardo@imaff.cfmac.csic.es} \affiliation{Departamento de
Matem\'aticas, Escuela Polit\'ecnica Superior, Universidad Carlos
III de Madrid, Avda. de la Universidad 30, 28911 Legan\'es, Spain}

\date{February 10, 2004}

\begin{abstract}
We give a detailed study of the asymptotic behavior of field
commutators for linearly polarized, cylindrically symmetric
gravitational waves in different physically relevant regimes. We
also discuss the necessary mathematical tools to carry out our
analysis. Field commutators are used here to analyze
microcausality, in particular the smearing of light cones owing to
quantum effects. We discuss in detail several issues related to
the semiclassical limit of quantum gravity, in the simplified
setting of the cylindrical symmetry reduction considered here. We
show, for example, that the small G behavior is not uniform in the
sense that its functional form depends on the causal relationship
between spacetime points. We consider several physical issues
relevant for this type of models such as the emergence of large
gravitational effects.
\end{abstract}

\pacs{04.60.Ds, 04.60.Kz, 04.62.+v.}

\maketitle


\section{\label{Intro}Introduction}

Among the many symmetry reductions of General Relativity that have
been considered in the past, linearly polarized cylindrical waves
(also known as Einstein-Rosen waves \cite{Einstein-Rosen,
Kuchar:1971xm}) have been the focus of intensive study. They
provide a model with an infinite number of degrees of freedom that
can be exactly solved in spite of the fact that it is non-linear.
This is not only true classically but also quantum mechanically,
and hence this system is a valuable tool to explore the physics
that may be found if a successful quantization of full general
relativity is ever achieved \cite{Ashtekar:1996bb,
Ashtekar:1996yk, Angulo:2000ad, BarberoG.:2003ye}.

One of the main reasons behind this success is the fact that the
physical Hamiltonian is a function of the free Hamiltonian of a
2+1 dimensional, axially symmetric, massless scalar field evolving
in an auxiliary Minkowskian background \cite{Ashtekar:1996yk,
Angulo:2000ad, Ashtekar:1997cm}. In a previous paper
\cite{BarberoG.:2003ye} we took advantage of this fact to study
the quantum corrections to the spacetime structure by considering
the commutator of this scalar field at different spacetime points.
As is well known, microcausality in quantum field theories can be
discussed by looking at the commutator of quantum fields (or
anticommutator in the case of fermions). In the standard examples
the microcausality requirement means that this (anti)commutator
must vanish for spatially separated spacetime points. A similar
argument can be made for vector fields, though issues of gauge
invariance change some of the conclusions; in particular if one
computes the commutator of the four-vector potential $A_{\mu}$ at
two spatially separated points, it may be different from zero in
some gauges, even though it is always true that the commutator of
gauge invariant objects is zero for such points.

The issue of gauge invariance in the context of cylindrical
gravitational waves has been discussed at length by Bi\v{c}\'ak
and collaborators \cite{Kouletsis:2003hj}. These authors show that
is is legitimate to use the Ashtekar-Pierri  gauge fixed action
\cite{Ashtekar:1996bb} written in terms of the scalar field that
encodes the physical information in this model to obtain gauge
invariant structures such as Dirac observables or the S matrix.
This justifies the computation of the type of objects --field
commutators-- that we will be considering here to extract
conclusions about the quantum structure of spacetime.

The particular problem that we will be concerned with in this
paper is the detailed study of the field commutator (given as a
certain integral) and, in particular, its limiting behavior when
the time and length separations are much larger than the natural
length scale of the problem --the Planck length--. In order to do
this the procedure of expanding the integrand of the field
commutator as a power series in the gravitational constant and
other asymptotic parameters is not useful. In fact it will be
necessary to adapt some methods developed for the asymptotic
analysis of integrals and get a consistent procedure to expand the
relevant objects as asymptotic series. It is possible to
understand this \textit{a posteriori} as a consequence of the fact
that some limiting behaviors (i.e. in $G$) of the field commutator
change in a non trivial way as the spacetime intervals go from
spacelike to timelike or when one of the spacetime points lies in
the symmetry axis. Also the functional dependence in some of these
parameters is highly non-polynomial. This is not what one would
expect to obtain in the familiar perturbative treatment of quantum
field theories (QFT's).

The paper is divided in two main sections: a physical discussion
of the behavior of the field commutator followed by a detailed
description of the asymptotic methods necessary to study the
different physically relevant regimes. Specifically, after this
introduction we will give the different asymptotic expansions for
all the relevant parameters (involving the $G\rightarrow0$ limit
and also the limits in which the difference in the time
coordinates or the radial coordinates go to infinity). Using them
we will discuss the main physical consequences of the quantization
of this model as far as microcausality is concerned. In this
respect it is particulary interesting to point out the existence
of a certain type of large quantum mechanical effects (in a sense
that will be made precise later) when one of the spacetime points
in the field commutator lies in the symmetry axis.

A technical issue that should be considered is the role of
regulators in the final physical results. As is well known
regulators are necessary to give sense to otherwise ill-defined
objects. They must be introduced, for example, to obtain a finite
norm vector by acting with the field operator on the vacuum state.
Cut-offs are a simple way to regulate amplitudes. It is
conceivable that physical regulators exist that restrict, for
example, the integration intervals for some physical objects (such
as the field commutators considered here) to finite real
intervals. However it is also possible that they are just a
convenient way to render some physical objects finite in such a
way that no footprint is left in the final results. This is the
philosophy that one has in mind in the usual renormalization
scheme where cut-offs are taken to infinity and disappear from
physical quantities.

The point of view of this paper is that the un-regulated objects
(integrals), when defined, are a good approximation to the
regulated ones. The asymptotic analysis of regulated commutators
and their relation with the un-regulated ones will be discussed
elsewhere.

In the usual perturbative QFT computations Green functions, matrix
elements, and similar objects are expanded as power series in the
coupling constants of the model with coefficients that are usually
written as regulated integrals. This is a necessary step because
it is usually not possible to write them in closed form. Here the
situation is different because it is possible to write down
expressions for the objects of interest (field commutators in the
present example) that depend on the coupling constants in a non
trivial way. This has the advantage of allowing us to use
approximation techniques specially adapted to their specific form
and much better suited to its study. It also permits to consider
some problems that may be difficult to tackle for the usual QFT's;
for example one can try to figure out if the asymptotic behavior
of some regulated object coincides with its asymptotic behavior
after the regulator is removed.

A detailed technical discussion of the techniques needed to
explore the relevant asymptotic regimes is presented in section
\ref{maths}. These techniques will be useful tools in order to
analyze other types of physical objects (such as S-matrix
elements) so they provide the necessary background for a
perturbative framework properly adapted to this model, this is why
we invest some time to study them here. We end the paper with a
summary of the main results, comments, perspectives for future
work, and several appendices. Numerical and algebraic computations
have been performed with the aid of
\emph{Mathematica}$\textregistered$.


\section{\label{comm} The field commutator}

We start with a brief description of the model and introduce our
conventions and notation. As is well known \cite{Romano:1996ep,
Kuchar:1971xm, Einstein-Rosen} Einstein-Rosen waves correspond to
topologically trivial spacetimes with two linearly independent,
commuting, spacelike, and hypersurface orthogonal Killing vector
fields. The metric in this case can be writen as
\begin{equation}
ds^2=e^{\gamma-\psi}(-dT^2+dR^2) +e^{-\psi}R^2d\theta^2+e^\psi
dZ^2,\label{metric}
\end{equation}
where we are using coordinates
$(T,R,\theta,Z),\,\,T\in\mathbb{R},\,\,R\in[0,\infty),\,\,
\theta\in[0,2\pi),\,\,Z\in\mathbb{R}$, and $\psi$ and $\gamma$ are
functions only of $R$ and $T$. The Einstein field equations for
this metric are very simple: the scalar field $\psi$ must satisfy
the wave equation for a massless, axially symmetric scalar field
in three dimensions
$$
\partial_T^2\psi-\partial_R^2\psi-\frac{1}{R}\partial_R\psi=0
$$
and the function $\gamma$ can be expressed in terms of this field
on the classical solutions \cite{Ashtekar:1996bb, Angulo:2000ad}
as
$$
\gamma(R)=\frac{1}{2}\int^R_0 d\bar{R} \,\bar{R}\, \left[
(\partial_T\psi)^2+(\partial_{\bar{R}}\psi)^2 \right].
$$
Throughout the paper we will use a system of units such that
$c=\hbar=1$ and define $G\equiv\hbar G_3$, where $G_3$ denotes the
gravitational constant per unit length in the direction of the
symmetry axis\footnote{Notice that in this system of units $G$ has
dimensions of length.}. The metric function $\gamma(R)$ admits a
simple physical interpretation: apart from a factor of $8G$ it is
the energy of the scalar field in a ball of radius $R$, and
$\gamma_{\infty}\equiv\lim_{R\rightarrow\infty}\gamma(R)$ the
energy of the whole two-dimensional flat space. Furthermore,
$\gamma_\infty/8G$ coincides with the Hamiltonian $H_0$ of the
system obtained by linearizing the metric
(\ref{metric})\cite{Ashtekar:1996bb, BarberoG.:2003ye}.

In order to arrive at a unit asymptotic timelike Killing vector
that allows us to introduce a physical notion of energy (per unit
length) it is convenient to use coordinates $(t,R,\theta,Z)$
defined by $T=e^{-\gamma_{\infty}/2}t$. In these coordinates the
metric takes the form \cite{Kuchar:1971xm, Romano:1996ep}
$$
ds^2=e^{\gamma-\psi}(-e^{-\gamma_\infty}dt^2+dR^2)
+e^{-\psi}R^2d\theta^2+e^\psi dZ^2.
$$
By choosing a metric function $\psi$ with a sufficiently fast
fall-off as $R\rightarrow \infty$, this metric describes
asymptotically flat spacetimes with a certain (non-zero) deficit
angle and such that $\partial_t$ is a unit timelike Killing vector
in the asymptotic region.

The Einstein field equations can be obtained from a Hamiltonian
action principle \cite{Ashtekar:1994ds, Varadarajan:1995hw,
Romano:1996ep} where the Hamiltonian $H$ is a function of that
corresponding to the free scalar field, $H_0$:
\begin{eqnarray*}
H=E(H_0)=\frac{1}{4G}(1-e^{-4GH_0}).
\end{eqnarray*}
In the following we will refer to $t$ as the physical time and to
$H$ as the physical Hamiltonian.

In terms of the $T$-time and imposing that $\psi$ be regular at
$R=0$ \cite{Ashtekar:1996bb}, the classical solutions for the
field $\psi$ can be written as
\begin{eqnarray*}
\psi(R,T)=\sqrt{4G}\int_0^\infty \!\!dk\, J_0(Rk)
\left[A(k)e^{-ikT} +A^\dagger(k)e^{ikT}\right]
\end{eqnarray*}
where $A(k)$ [and its complex conjugate $A^\dagger(k)$] are
determined by the initial conditions\footnote{Throughout the paper
improper integrals are understood in the Riemann sense
$\int_a^{\infty}=\lim_{A\rightarrow\infty}\int_a^A$.}. The free
Hamiltonian $H_0$ can be written now as
\begin{eqnarray*}
\gamma_{\infty}=H_0=\int_0^{\infty}dk\,kA^{\dagger}(k)A(k).
\end{eqnarray*}
Using this expression, we can obtain the $t$-evolution of the
field
\begin{eqnarray*}
\psi_E(R,t)= \sqrt{4G}\int_0^\infty\!\! dk\,
J_0(Rk)\left[A(k)e^{-ikte^{-\gamma_\infty/2}}
\!+\!A^\dagger(k)e^{ikte^{-\gamma_\infty/2}}\right].\;
\end{eqnarray*}

The quantization can be carried out by following the usual steps.
We introduce a Fock space in which $\hat{\psi}(R,0)$, the quantum
counterpart of $\psi(R,0)$, is an operator-valued distribution
\cite{Reed:1975uy}. Its action is determined by those of
$\hat{A}(k)$ and $\hat{A}^\dagger(k)$, the annihilation and
creation operators with non-vanishing commutators given by
\begin{eqnarray*}
\left[\hat{A}(k_1),\hat{A}^\dagger(k_2)\right]=\delta(k_1,k_2).
\end{eqnarray*}
Explicitly,
\begin{eqnarray*}
\hat{\psi}(R,0)\!=\!\hat{\psi}_E(R,0)\!=\!\sqrt{4G}\int_0^\infty
\!\!\!dk\,
J_0(Rk)\!\left[\hat{A}(k)\!+\!\hat{A}^\dagger(k)\right]\!.
\label{psi0}
\end{eqnarray*}
Evolution in $T$ is given by the unitary operator
$\hat{U}_0(T)=\exp(-iT\hat{H}_0)$ where
\begin{eqnarray*}\label{qham0}
\hat{H}_0=\int^{\infty}_0dk\,k\,\hat{A}^{\dagger}(k)\hat{A}(k)
\end{eqnarray*}
is the quantum Hamiltonian operator of a three dimensional,
axially symmetric scalar field. The quantum scalar field in the
Heisenberg picture is hence given by
\begin{eqnarray}
\hspace*{-.3cm}\hat{\psi}(R,T)=
\hat{U}^\dagger_0(T)\hat{\psi}(R,0)\hat{U}_0(T)=\sqrt{4G}\int_0^\infty\!\!
dk\,
J_0(Rk)\!\left[\hat{A}(k)e^{-ikT}\!+\!\hat{A}^\dagger(k)e^{ikT}\right]
\!.\hspace*{.3cm}\nonumber
\end{eqnarray}
If we choose the physical time $t$ to define the evolution in our
model, the quantum Hamiltonian is
$\hat{H}=E(\hat{H}_0)=\frac{1}{4G}(1-e^{-4G\hat{H}_0})$ and
unitary evolution is given by $\hat{U}(t)=\exp(-it\hat{H})$. With
this time evolution the annihilation and creation operators in the
Heisenberg picture are
\begin{eqnarray}
\hat{A}_E(k,t)\!\!&\equiv&\!\!\hat{U}^\dagger(t)\hat{A}(k)\hat{U}(t)=
\exp\!\left[-itE(k)e^{-4G\hat{H}_0}\right]\!\hat{A}(k),\nonumber\\
\hat{A}_E^\dagger(k,t)\!\!&=&\!\hat{A}^\dagger(k)\,\exp\!\left[itE(k)
e^{-4G\hat{H}_0}\right],\nonumber
\end{eqnarray}
where $E(k)=\frac{1}{4G}(1-e^{-4Gk})$, and the field operator
evolved with the physical Hamiltonian [that we denote as
$\hat{\psi}_E(R,t)$] is given by
\begin{eqnarray*}
\hat{\psi}_E(R,t)=\sqrt{4G}\int_0^\infty\!\!\!dk
\,J_0(Rk)\left[\hat{A}_E(k,t)+\hat{A}_E^\dagger(k,t)\right].
\end{eqnarray*}
The field commutator $\big[\hat{\psi}_E(R_1,t_1),
\hat{\psi}_E(R_2,t_2)\big]$ can be computed from these expressions
\cite{BarberoG.:2003ye}. One of its interesting features is the
fact that it is not proportional to the identity as in the case of
free theories (or if we consider the quantum evolution defined by
the Hamiltonian $H_0$ of the linearized model) but is a
non-diagonal operator in the chosen basis. We are hence led to
consider its matrix elements. We will concentrate here on the most
relevant of these elements (at least from the perspective of the
microcausality of the classical background of the model), namely
the vacuum expectation value, which is explicitly given by
\begin{eqnarray}
\frac{1}{8iG}\langle 0|\,
[\hat{\psi}_E(R_1,t_1),\hat{\psi}_E(R_2,t_2)]\,|0\rangle=
\int_0^{\infty}\!\!\!\!\!\!dk\,J_0(R_1 k)J_0(R_2
k)\sin\left[\frac{t_2-t_1}{4G}(1\!-\!e^{-4Gk})\right]\!.
\hspace*{.4cm}&&\label{integral}
\end{eqnarray}
As we can see it depends on the time coordinates through their
difference $t_2-t_1$ (which we will assume to be, e.g., positive)
and depends symmetrically on $R_1$ and $R_2$. The functional
dependence in $G$ is less trivial, a fact that will require
especial attention when studying the limit in which the relevant
lengths and time differences are much larger than the Planck
length. For comparison purposes we remember that
\begin{eqnarray}
\frac{1}{8iG}\langle 0|\,
[\hat{\psi}(R_1,T_1),\hat{\psi}(R_2,T_2)]\,|0\rangle=
\int_0^{\infty}\!\!\!\!\!\!dk\,J_0(R_1 k)J_0(R_2
k)\sin\left[k(T_2-T_1)\right]\!.
\hspace*{.4cm}&&\label{freeintegral}
\end{eqnarray}

In the following it will be convenient to refer the dimensional
parameters of the integral to another length scale that we choose
as $R_1$. We hence define $R_2=\rho R_1$, $t_2-t_1=\tau R_1$,
$\lambda=R_1/4G$ and rewrite (\ref{integral}) as
\begin{equation}
\frac{1}{8iG}\langle 0|\,
[\hat{\psi}_E(R_1,t_1),\hat{\psi}_E(R_2,t_2)]\,|0\rangle=
\frac{\lambda}{R_1}\Im\bigg\{\int_0^{\infty}dqJ_0(\lambda
q)J_0(\rho \lambda
q)e^{i\tau\lambda(1-e^{-q})}\bigg\}\label{newintegral}
\end{equation}
after introducing the new variable $k=q/4G$. Here $\Im$ denotes
the imaginary part.

\subsection{\label{r} Asymptotic behavior in $\rho$: smearing of light cones}

The integral (\ref{newintegral}) can be written as a standard
$h$-transform \cite{Handels}, with asymptotic parameter $\rho$, by
the change of variables $t=q\lambda$,
\begin{equation}
\frac{1}{R_1}\Im\bigg\{\int_0^{\infty}\!\!dt\,J_0(\rho
t)J_0(t)e^{i\tau\lambda(1-e^{-t/\lambda})}\bigg\}.\label{r-int-inf}
\end{equation}
In this case a straightforward Mellin transform analysis
\cite{Handels} (discussed in section \ref{maths}) gives the
following asymptotic behavior in the $\rho\rightarrow \infty$
limit
\begin{eqnarray}
& &
\frac{1}{R_1}\Im\bigg\{\frac{1}{\rho}+\frac{1}{\rho^3}\Big(\frac{1}{4}+\frac{\tau^2}{2}+
\frac{i\tau}{2\lambda}\Big)+\frac{9}{\rho^5}\Big(\frac{1}{64}-
\frac{i\tau}{24\lambda^3}+\frac{i\tau}{8\lambda}+\frac{\tau^2}{8}+
\frac{i\tau^2}{4\lambda}-\frac{7\tau^2}{24\lambda^2}+\frac{\tau^4}{24}\Big)+
O\Big[\frac{1}{\rho^7}\Big]\bigg\}\nonumber\\
& &=\frac{1}{R_1}\bigg\{\frac{\tau}{2\lambda
\rho^3}+\frac{1}{\rho^5}\Big(\frac{9\tau}{8\lambda}-\frac{3\tau}{8\lambda^3}+
\frac{9\tau^2}{4\lambda}\Big)+O\Big[\frac{1}{\rho^7}\Big]\bigg\}.\label{rasympt}
\end{eqnarray}
The asymptotic behavior in the $\rho\rightarrow\infty$ limit gives
a precise and quantitative description of the smearing of
(cylindrically symmetric) light cones because it shows, for
example, that for fixed values of $R_1$ and $(t_2-t_1)$ the
expectation value of the commutator (over $8iG$) falls off for
large values of $R_2$ as
$$
\frac{2G(t_2-t_1)}{R_2^3}+O\bigg[\frac{1}{R_2^5}\bigg].
$$
This means that even for large spatial separations the scalar
field that encodes the gravitational field does not commute with
itself.

In the $\rho\rightarrow 0^+$ limit, on the other hand, we get
\begin{equation}
\frac{1}{R_1}\Im\bigg\{\int_0^{\infty}dt
J_0(t)e^{i\tau\lambda(1-e^{-t/\lambda})}\bigg\}+O(\rho).\label{r-int-0}
\end{equation}
This shows that the commutator is a continuous function of $\rho$
in $\rho=0$, a fact that will be important in the analysis of the
semiclassical limit.

\subsection{\label{tau} Asymptotic behavior in $\tau$: large quantum gravity effects}

The integral in (\ref{r-int-inf}) has the convenient form of a
$h$-transform if the asymptotic parameter is chosen to be $\rho$;
however this is no longer true if the asymptotic parameter is
$\tau$ (which corresponds to considering large separations in the
time coordinates). This introduces some mathematical difficulties
in the asymptotic analysis that will be discussed later.

In this case one has to perform separate analyses for $\rho=0$ and
$\rho\neq0$. In $\rho=0$ one finds that the asymptotic behavior
when $\tau\rightarrow\infty$ is given by
\begin{equation}
\frac{1}{R_1}\bigg[\frac{\lambda}{2\pi\log\tau}\bigg]^{1/2}
       \Im\bigg\{e^{i[\frac{\pi}{4}+\tau\lambda-\lambda\log(\tau\lambda)]}
       e^{\frac{\pi}{2}\lambda}\Gamma(i\lambda)+
       e^{-i[\frac{\pi}{4}-\tau\lambda-\lambda\log(\tau\lambda)]}
       e^{-\frac{\pi}{2}\lambda}\Gamma(-i\lambda)\bigg\}+
       O\bigg[\frac{1}{(\log\tau)^{3/2}}\bigg],\label{tauasimpr0}
\end{equation}
whereas for $\rho\neq0$ we find
\begin{eqnarray}
&&\frac{1}{2 \pi R_1 \sqrt{\rho}
       \log\tau}\times\nonumber\\
       &&\Im\bigg\{e^{i[\frac{\pi}{2}+\tau\lambda-\lambda(1+\rho)\log(\tau\lambda)]}
       e^{\frac{\pi}{2}\lambda(1+\rho)}
       \Gamma[i\lambda(1+\rho)]+e^{-i[\frac{\pi}{2}-\tau\lambda-\lambda(1+\rho)\log(\tau\lambda)]}
       e^{-\frac{\pi}{2}\lambda(1+\rho)}
       \Gamma[-i\lambda(1+\rho)]\hspace{6mm}\label{tauasimprneq0}\\
       &&+e^{i[\tau\lambda-\lambda(1-\rho)\log(\tau\lambda)]}
       e^{\frac{\pi}{2}\lambda(1-\rho)}
       \Gamma[i\lambda(1-\rho)]+e^{i[\tau\lambda-\lambda(\rho-1)\log(\tau\lambda)]}
       e^{\frac{\pi}{2}\lambda(\rho-1)}
       \Gamma[i\lambda(\rho-1)]\bigg\}+O
       \bigg[\frac{1}{(\log\tau)^2}\bigg].\nonumber
\end{eqnarray}
The most striking feature of these expressions is the unusual
dependence on the asymptotic parameter $\tau$; in fact the
dependence on inverse powers of logarithms (especially on the
inverse square root of $\log\tau$) cannot be obtained by direct
application of the usual asymptotic expressions derived by Mellin
transform techniques. It is also remarkable how slowly the
commutator decays in $\tau$ --in particular in the axis $\rho=0$--
a fact that is suggestive of the large quantum gravity effects
discussed by Ashtekar in \cite{Ashtekar:1996yk}. Outside the axis
the decay is quicker but still rather slow. A consequence of the
different asymptotic behaviors in $\tau$ for $\rho=0$ and
$\rho\neq0$ is the impossibility to recover (\ref{tauasimpr0}) as
the limit when $\rho\rightarrow0$ of (\ref{tauasimprneq0}). It is
interesting to point out that the frequency of the oscillations of
the commutator in the $\tau$ parameter is controlled by the value
of $\lambda$ (proportional to the inverse of $G$) in such a way
that although the amplitude of the oscillations decays very slowly
they will average to the value of the free commutator on scales
much larger than the Planck length.

The $\tau\rightarrow0$ limit is simpler to analyze. Actually we
find that the series obtained by expanding
$e^{i\tau\lambda(1-e^{-t/\lambda})}$ as a power series in
$e^{-t/\lambda}$, exchanging integration and infinite sum, and
computing the resulting integrals gives an expansion that
converges to the value of the commutator.

\subsection{\label{lambda} Asymptotic behavior in $\lambda$: the semiclassical limit}

The possibility of studying the asymptotic behaviors in $\rho$ and
$\tau$ by using the powerful mathematical tools provided by the
theory of Mellin transforms relies on the fact that these
integrals can be written as $h$-transforms of the type mentioned
above. However this is no longer possible if one wants to study
the limit $\lambda\rightarrow\infty$ (corresponding to the
situation when $R_1$ is much larger than the Planck length)
because of the particular dependence of these integrals on
$\lambda$. There is, however, a way out of this if one is willing
to abandon Mellin transforms: writing the integral in
(\ref{newintegral}) as a multiple integral by introducing the
integral representation of the Bessel functions $J_n$
($n=0,\,1,\,\ldots$),
\begin{eqnarray*}
J_n(z)=\frac{1}{2\pi i}\oint_{\gamma}
\frac{dt}{t^{n+1}}e^{\frac{z}{2}(t-\frac{\scriptstyle
1}{\scriptstyle t})},
\end{eqnarray*}
where $\gamma$ is a closed, positively oriented, simple path in
the complex plane surrounding the origin. By doing this the right
hand side (r.h.s.) of (\ref{newintegral}) can be rewritten in the
following form
\begin{equation}
-\frac{1}{R_1}\Im\Big\{\frac{\lambda
e^{i\tau\lambda}}{4\pi^2}\int_0^{\infty}dq\oint_{\gamma_1}
dt_1\oint_{\gamma_2}dt_2\frac{1}{t_1t_2}
e^{\lambda\big[\frac{\scriptstyle
q}{\scriptstyle2}(t_1-\frac{\scriptstyle1}{\scriptstyle
t_1})+\frac{\scriptstyle \rho
q}{\scriptstyle2}(t_2-\frac{\scriptstyle1}{\scriptstyle
t_2})-i\tau e^{-q}\big]}\Big\}.\label{trintegral}
\end{equation}
Although the contours $\gamma_1$ and $\gamma_2$ may be different
in principle we will take them equal in practice. Besides, we will
see in the next section that it is convenient to choose them in a
specific way.

\begin{figure}
\includegraphics[width=8.5cm]{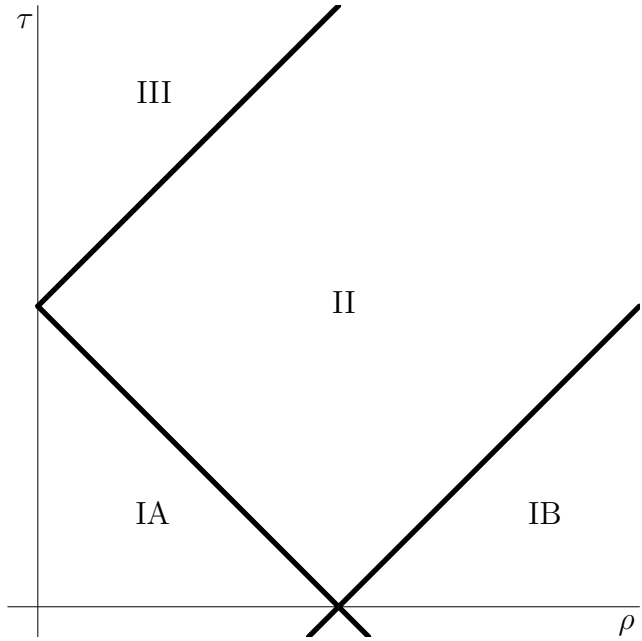}
\caption{Regions in the $(\rho,\tau)$ plane used in the discussion
of the $\lambda$ asymptotics. Region I corresponds to
$0<\tau<|\rho-1|$, region II to $|\rho-1|<\tau<\rho+1$, and region
III to $\rho+1<\tau$.} \label{fig1:regions}
\end{figure}

As in the $\tau$ case one has to perform separate asymptotic
analyses for $\rho=0$ and $\rho\neq0$. In order to proceed it is
necessary to divide the $(\rho,\,\tau)$ plane into the same
regions that appear in the discussion of the ``free" commutator
(\ref{freeintegral}) which corresponds to the evolution dictated
by the Hamiltonian $H_0$ \cite{BarberoG.:2003ye} (see fig.
\ref{fig1:regions}).

\begin{figure}
\hspace{-0.5cm}\includegraphics[width=16cm]{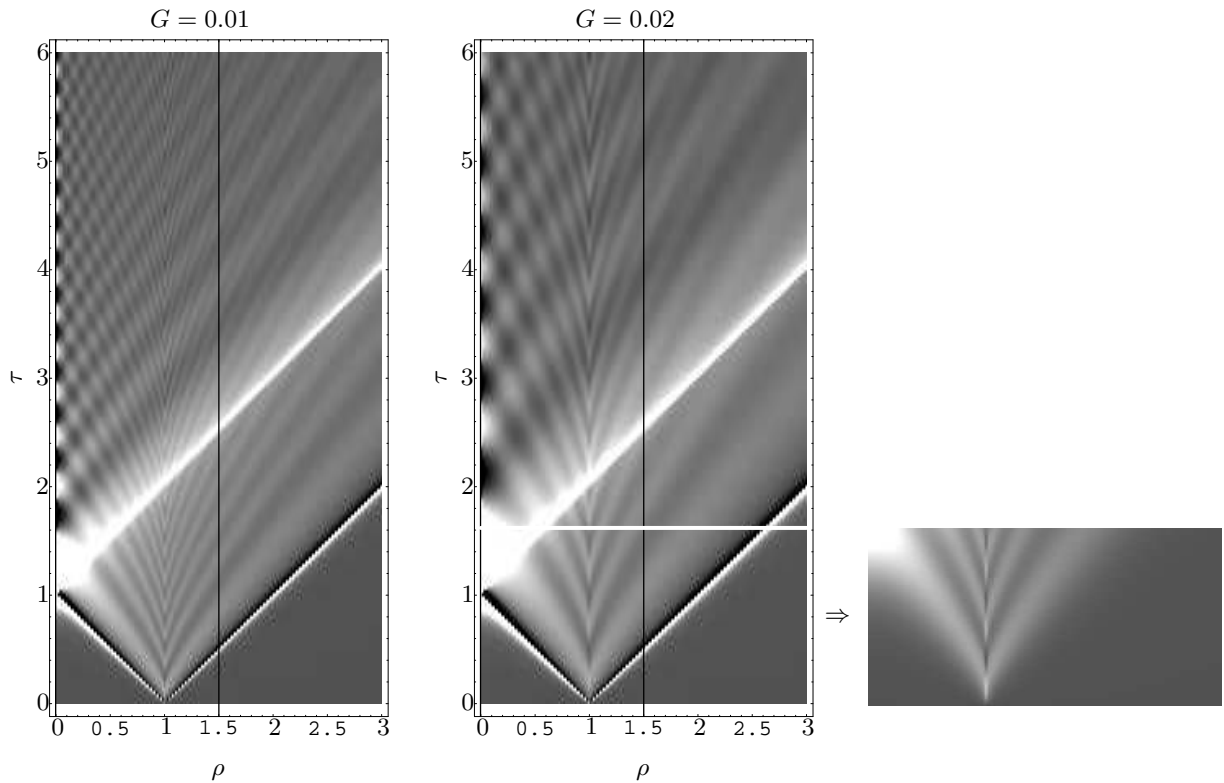}
\caption{Density plot of the asymptotic approximation in $\lambda$
for the field commutator (over $8iG$) as a function of
$(\rho,\tau)$ for $G=0.01$ and $G=0.02$. Notice the singular
behavior of the approximation in the boundaries between the
regions I, II, III, and on the axis. For comparison we show part
of the plot for $G=0.02$ obtained with the series expansion
(\ref{series}). Sections of these plots for $\rho=0$ and
$\rho=1.5$ with $G=0.02$ are shown in figs. \ref{fig3:commutator}
and \ref{fig4:commutator}.} \label{fig2:commutator}
\end{figure}

\begin{figure}
\hspace{0cm}\includegraphics[width=16.5cm]{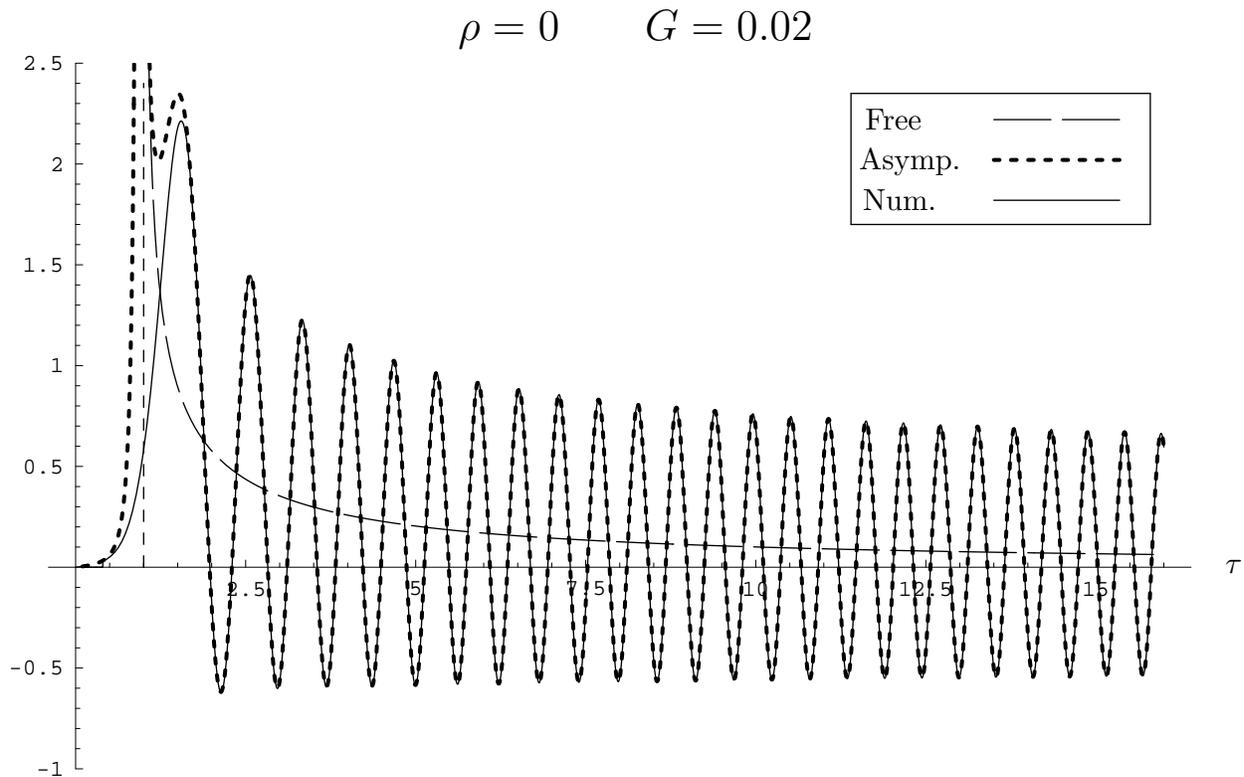}
\caption{Asymptotic approximation in $\lambda$ for the field
commutator (over $8iG$) as a function of $\tau$ for $\rho=0$ and
$G=0.02$. We plot, for comparison, the commutator corresponding to
the free Hamiltonian discussed in \cite{BarberoG.:2003ye}. The
solid line corresponds to a numerical computation of
(\ref{newintegral}). As can be seen the asymptotic approximation
is very good for most of the values of $\tau$.}
\label{fig3:commutator}
\end{figure}

\begin{figure}
\hspace{0cm}\includegraphics[width=16.5cm]{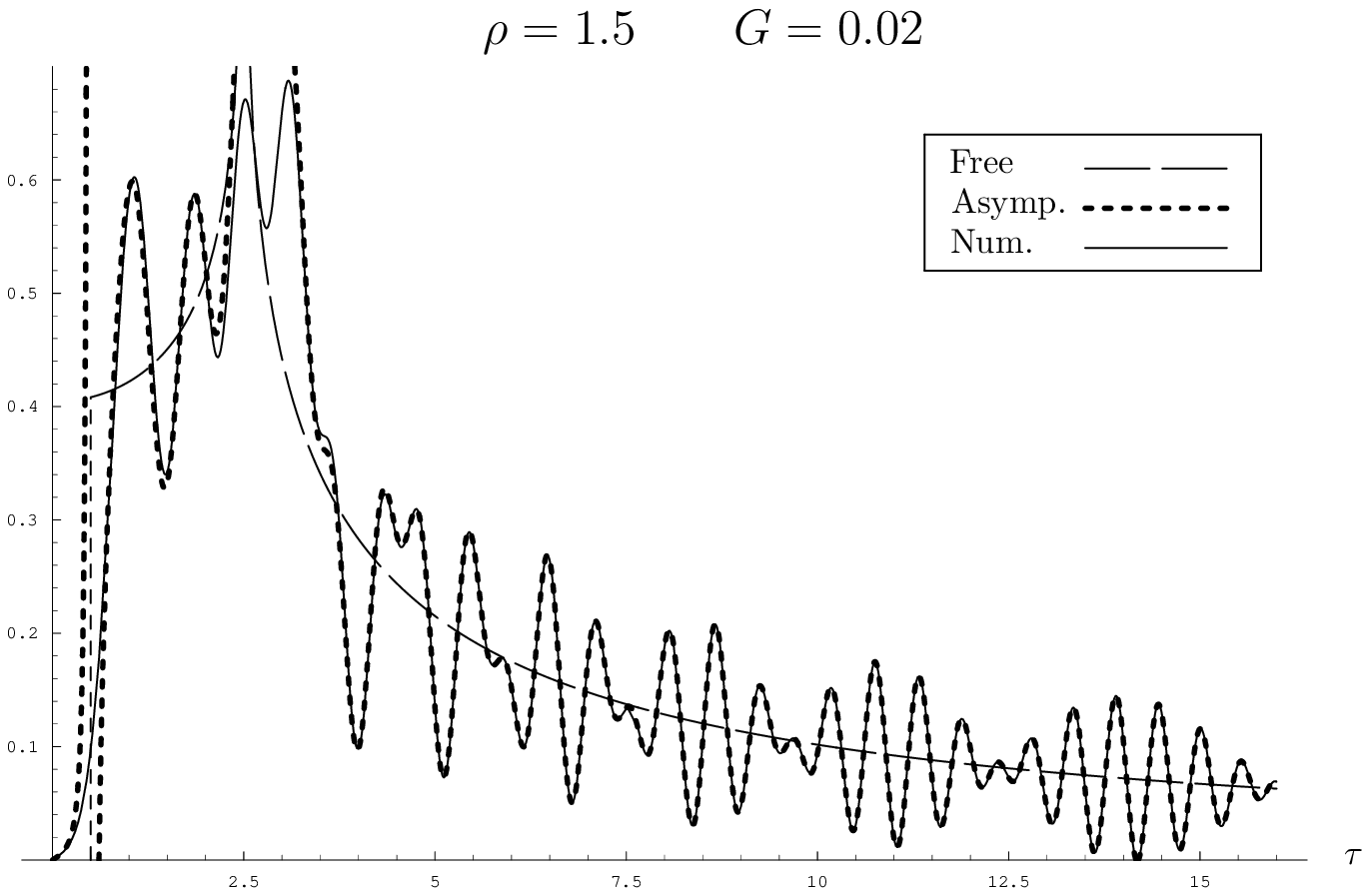}
\caption{Asymptotic approximation  in $\lambda$ for the field
commutator (over $8iG$) as a function of $\tau$ for $\rho=1.5$ and
$G=0.02$. We plot, as before, the contribution corresponding to
the free Hamiltonian and a numerical computation of the
commutator.} \label{fig4:commutator}
\end{figure}

In $\rho=0$ one finds that the asymptotic behavior when
$\lambda\rightarrow\infty$ is given by
\begin{eqnarray}
\frac{1}{R_1}\Im\bigg\{\frac{i}{\sqrt{\tau^2-1}}+
\frac{e^{i\lambda[\tau-\log\tau-1]}}{\sqrt{\log\tau}}\bigg\}+
O\bigg[\frac{1}{\lambda}\bigg]\hspace{1cm}\rm{\tau> 1},&&\label{lambdasimpt1}\\
\frac{1}{R_1}\frac{\tau(1+2\tau^2)}{2\lambda(1-\tau^2)^{5/2}}+
O\bigg[\frac{1}{\lambda^2}\bigg]\hspace{3.2cm}\rm{\tau<
1}.\label{lambdasimpt2}&&
\end{eqnarray}
If $\rho\neq0$ the asymptotic behavior in the different regions
shown in fig. \ref{fig1:regions} is the following:

\bigskip
\noindent Regions IA and IB
\begin{eqnarray}
&&\frac{\tau}{2\pi R_1\lambda}\left\{
\frac{2[1+\rho^4+2\tau^2-3\tau^4+2\rho^2(\tau^2-1)]\sqrt{(1+\rho)^2-\tau^2}}
{(1+\rho-\tau)^2(1-\rho+\tau)^2(-1+\rho+\tau)^2(1+\rho+\tau)^2}
E\left(\sqrt{\frac{4\rho}{(1+\rho)^2-\tau^2}}\right)\right.\nonumber\\
&&\left.-\frac{2\tau^2}
{[\rho^4+(\tau^2-1)^2-2(1+\tau^2)\rho^2]\sqrt{(1+\rho)^2-\tau^2}}
K\left(\sqrt{\frac{4\rho}{(1+\rho)^2-\tau^2}}\right)\right\}+
O\bigg[\frac{1}{\lambda^{2}}\bigg].\label{IAIB}
\end{eqnarray}
\noindent Region II
\begin{eqnarray}
&&\frac{1}{\pi
R_1\sqrt{\rho}}K\left(\sqrt{\frac{\tau^2-(\rho-1)^2}{4\rho}}\right)+
\frac{1}{R_1}\Im\bigg\{\frac{e^{-i\frac{\pi}{4}}
e^{i\lambda[\tau+|\rho-1|(1+\log\frac{\tau}{|\rho-1|})]}}
{\sqrt{2\pi\lambda \rho|1-\rho|}\log\frac{\tau}{|1-\rho|}}\bigg\}+
O\bigg[\frac{1}{\lambda^{3/2}}\bigg].\label{II}
\end{eqnarray}
\noindent Region III
\begin{eqnarray}
&&\frac{2}{\pi R_1}\frac{1}{\sqrt{\tau^2-(1-\rho)^2}}
K\left(\sqrt{\frac{4\rho}{\tau^2-(1-\rho)^2}}\right)\hspace{5.5cm}\nonumber\\
&&\hspace{2cm}+\frac{1}{R_1} \Im\bigg\{\frac{e^{-i\frac{\pi}{4}}
e^{i\lambda[\tau-|\rho-1|(1+\log\frac{\tau}{|\rho-1|})]}}
{\sqrt{2\pi\lambda \rho|1-\rho|}\log\frac{\tau}{|1-\rho|}}+
\frac{e^{i\frac{\pi}{4}}e^{i\lambda[\tau+(\rho+1)(\log\frac{1+\rho}{\tau}-1)]}}
{\sqrt{2\pi\lambda \rho(1+\rho)}\log\frac{\tau}{1+\rho}}\bigg\}+
O\bigg[\frac{1}{\lambda^{3/2}}\bigg].\label{III}
\end{eqnarray}
Here $K(k)$ and $E(k)$ denote the complete elliptic integrals of
the first and second kind respectively, defined by
$$
K(k)\equiv\int_0^{\frac{\pi}{2}}\frac{d\theta}{\sqrt{1-k^2\sin^2\theta}},\quad\quad
E(k)\equiv\int_0^{\frac{\pi}{2}}d\theta\sqrt{1-k^2\sin^2\theta}.
$$
Some comments are in order now. The first is that the $\lambda$
independent terms in the above expressions correspond to the
commutator obtained from the free Hamiltonian $H_0$ both in the
axis and outside the axis. The remaining terms (except when
$\rho=0$ and $\tau>1$) are corrections to this free commutator
that fall off to zero as $\lambda\rightarrow\infty$, and have an
additional, non-polynomial dependence in $1/\lambda$. Since the
free commutator defines a characteristic light cone structure
these terms are responsible for the smearing of the light cones in
this model. It is worthwhile pointing out that the asymptotic
behavior in $\lambda$ is different in regions I, II, III, and in
the axis --this is the reason that explains the appearance of
singularities in the borders between adjacent regions\footnote{As
shown in \cite{BarberoG.:2003ye} the only real singularity that
appears now corresponds to $\rho=1$.}-- and it has a
non-polynomial dependence on $1/\lambda$. This kind of behavior
cannot possibly appear in an ordinary perturbative QFT where the
relevant objects (propagators, Green's functions, and so on) are
expanded as power series in the coupling constants. One of the
novel features of the approach that we follow in this paper is
that by using more sophisticated approximation techniques, and
taking advantage of the fact that we have closed explicit
expressions for the objects of interest, we are able to extract
such non trivial behaviors. In the axis we notice that for
$\tau>1$ there is a $\lambda$-independent contribution that
corresponds to the free commutator and an oscillating contribution
with a frequency that depends on $\lambda$ (this term is similar
to the one obtained in the asymptotic analysis for $\tau$). For
large values of $\lambda$ this term oscillates very fast and
averages to zero. For $\tau<1$ we get a correction to the free
commutator (which is zero in region I) that goes to zero as
$1/\lambda$. Outside the axis we see that the corrections to the
free commutator fall-off as $1/\lambda$ in region I --outside the
light cone--, but only as $1/\sqrt{\lambda}$ (with the additional
non-polynomial terms) in regions II and III. The presence of two
oscillating terms in region III produces some interference effects
that manifest themselves as a checkered pattern in plots of the
commutator --especially close to the axis-- which suggests a
division of spacetime into cells of a size governed by the value
of the gravitational constant (through $\lambda$). These are shown
in fig. \ref{fig2:commutator} where we plot the free commutator
over $8iG$ plus the first asymptotic correction, given by the
above expressions in each of the regions. For comparison we also
display a plot constructed with the power series representation
obtained in \cite{BarberoG.:2003ye}:
\begin{eqnarray}
\frac{1}{\pi
R_1\sqrt{\rho}}\left\{\!\sin(\lambda\tau)\!\!\sum_{n=0}^{\infty}
\!\frac{\!(-1)^n(\lambda\tau)^{2n}\!\!}
{(2n)!}Q_{-\frac{1}{2}}\left[\sigma_{2n}(\rho)\right]-\!\cos(\lambda\tau)\!\sum_{n=0}^{\infty}
\!\frac{\!(-1)^n(\lambda\tau)^{2n+1}\!\!\!}
{(2n+1)!}Q_{-\frac{1}{2}}\!\left[\sigma_{2n+1}(\rho)\right]\!\right\},\label{series}
\hspace*{.3cm}&&
\end{eqnarray}
where
\begin{eqnarray*}
\sigma_n(\rho)=\frac{n^2+\lambda^2(1+\rho^2)}{2\rho\lambda^2}
\end{eqnarray*}
and $Q_{-\frac{1}{2}}(x)=\pi
F\left(\frac{3}{4},\frac{1}{4};1;\frac{1}{x^2}\right)/\sqrt{2x}$
[for $x>1$] is the associated Legendre function of the second kind
\cite{Gradshteyn}. As can be seen, the result of the asymptotic
approximation matches that obtained with the power series
expansion (\ref{series})\footnote{The very slow convergence of
this oscillating series for large values of $\tau$ or large values
of $\lambda$ makes it impractical for numerical computations. This
is the reason why we only give the lower portion of the plot in
fig. \ref{fig2:commutator}.}. In figs. \ref{fig3:commutator} and
\ref{fig4:commutator} we also plot the commutator as a function of
$\tau$ for $\rho=0$ and a value of $\rho$ different from zero. It
is interesting to point out that, as $\tau$ grows, a distinctive
beating pattern appears due to the interference of terms in the
asymptotic expressions for region III mentioned above.

Comparison with the results of numerical integration confirm the
accuracy of the approximation provided by the asymptotic
expressions, as long as one is far enough from the boundary
between regions. This is shown in figs. \ref{fig3:commutator} and
\ref{fig4:commutator}, where we compare the asymptotic
approximation with the result obtained by numerically computing
the field commutator at $\rho=0$ and $\rho=1.5$.

The limit $\lambda\rightarrow0$ is directly obtained by truncating
(\ref{series}) to the desired number of terms.


\section{\label{maths} Asymptotic analysis: Mathematical details}

We discuss in detail here the techniques used to obtain the
asymptotic expansions discussed in the previous section. Even
though we will be mostly employing standard techniques it is
nonetheless necessary to adapt them to the different parameters
that appear in the problem, ($\rho$, $\tau$, $\lambda$). Before
studying the asymptotic expansions we present some results on
Mellin transforms and explain the basics of their use to obtain
the asymptotic behavior of integrals.

Whenever possible it will prove useful to write the integrals
under consideration as $h$-transforms, i.e. objects of the type
$$
H[f; \zeta]:=\int_0^{\infty}h(\zeta t)f(t) dt
$$
where $\zeta$ is the asymptotic parameter (that in practice
approaches either 0 or $\infty$) and $f$ and $h$ are locally
integrable functions in $(0,\infty)$ [functions that are
integrable in every closed interval in $(0,\infty)$]. This will
allow us, in many cases, to employ standard asymptotic techniques
(see e.g. \cite{Handels}) based on Mellin transforms\footnote{The
Mellin transform of a locally integrable function $f$ in
$(0,\infty)$ is given by $\int_0^{\infty}dt\,\, t^{z-1}f(t)$. It
is a holomorphic function defined on the strip
$\alpha<\Re(z)<\beta$ of the complex plane where the integral is
absolutely convergent.} that can be summarized in the following

\bigskip
\noindent\textbf{theorem:} \textit{Let
$h:(0,\infty)\mapsto\mathbb{C}$ and
$f:(0,\infty)\mapsto\mathbb{C}$ be locally integrable functions.
Let us assume that they respectively have the following asymptotic
expansions when $t\rightarrow\infty$ and $t\rightarrow0^+$:
\begin{equation}
h(t)\sim e^{-d
t^{\nu}}\sum_{m=0}^{\infty}\sum_{n=0}^{N(m)}c_{mn}t^{-r_m}(\log
t)^n\label{th1}
\end{equation}
{\rm[}$\Re(d)\geq0$, $\nu>0$, $\Re(r_m)$ grows monotonically to
$\infty$, and $N(m)>0$ is a finite integer for each value of
$m${\rm]} and
\begin{equation}
f(t)\sim e^{-q
t^{\mu}}\sum_{m=0}^{\infty}\sum_{n=0}^{\bar{N}(m)}p_{mn}t^{a_m}(\log
t)^n\label{th2}
\end{equation}
{\rm[}$\Re(q)\geq0$, $\mu>0$, $\Re(a_m)$ grows monotonically to
$\infty$ and $\bar{N}(m)>0$ is a finite integer for each value of
$m${\rm]}.}

\textit{Let $M[h; z]$ and $M[f;z]$ be the Mellin transforms of $h$
and $f$ (in the generalized sense, see \cite{Handels} for
details), holomorphic in the strips $\alpha<\Re(z)<\beta$ and
$\gamma<\Re(z)<\delta$ that we assume overlapping. Let us suppose
also that the Parseval identity holds:\footnote{Conditions for
this identity to hold can be found in \cite{Handels}.}
\begin{equation}
H[f;\zeta]=\frac{1}{2\pi i}\int_{r-i\infty}^{r+i\infty}\!\!dz
\zeta^{-z}M[h;z]M[f;1-z]\equiv\frac{1}{2\pi
i}\int_{r-i\infty}^{r+i\infty}\!\!dz\zeta^{-z}G(z) \label{th3}
\end{equation}
with $r$ in the intersection of the holomorphicity strips. If
$\lim_{|y|\rightarrow\infty}G(x+iy)=0$ for each $x\in[r,R]$,
($R\in\mathbb{R}$) and $\int_{-\infty}^{\infty}dy|G(R+iy)|$ is
finite \textbf{then}
$$
H[f;\zeta]\sim -\!\!\!\!\!\!\sum_{r<\Re(z)<R}{\rm
res}(\zeta^{-z}M[h;z]M[f;1-z])
$$
is a finite asymptotic expansion of $H[f;\zeta]$ as
$\zeta\rightarrow\infty$ with respect to the asymptotic sequence
$\Phi_{j,m}(\zeta)=\{\zeta^{-\alpha_j}(\log\zeta)^{n_j-m}\}$,
$m=0,\,1,\ldots,n$, $j=0,\,1,\ldots$ with error $O(\zeta^{-R})$.
If the previous hypotheses hold for arbitrary values of $R$ we get
an asymptotic expansion with an infinite number of terms.}

In the following we will only need to consider the case when
$d\neq0$ and $q=0$. The result of the previous theorem can be
written as
\begin{equation}
H[f;\zeta]\sim\sum_{m=0}^{\infty}
\zeta^{-1-a_m}\sum_{n=0}^{\bar{N}(m)}p_{mn}\sum_{j=0}^{n}
\left(\begin{array}{c}n\\j\end{array}\right)(-\log\zeta)^jM^{(n-j)}[h;z]\Big|_{z=1+a_m}
\label{th4}
\end{equation}
with $M^{(k)}[h;z]$ denoting the $k^{\rm{th}}$ derivative of the
Mellin transform with respect to to $z$.

\subsection{\label{r-math} Asymptotic behavior in $\rho$}

We write the integral that gives the field commutator in the form
(\ref{r-int-inf}) and choose $\rho$ as the asymptotic parameter.
In order to apply the previous theorem we need the asymptotic
expansion of $f(t)=J_0(t) e^{i\tau \lambda (1-e^{-t/\lambda})}$ as
$t\rightarrow0^+$ and the Mellin transform of $h(t)=J_0(t)$. From
the first we get $a_k=k$, ($k=0,\,1,\ldots$),
$p_{00}=1,\,p_{10}=i\tau$, \emph{etc.} The Mellin transform of $h$
is given by
$$
M[h;z]=\frac{2^{z-1}\Gamma(z/2)}{\Gamma(1-z/2)}.
$$
Since $M[h;k]=0$ for even values of $k$, we get the asymptotic
expansion in odd powers of $1/\rho$ given in (\ref{rasympt}).

In order to obtain the asymptotic expansion of the commutator when
$\rho\rightarrow0^+$ we change variables in (\ref{r-int-inf})
according to $s=\rho t$,
$$
\frac{1}{\rho R_1}\Im\bigg\{\int_0^{\infty}
ds\,J_0(s)J_0\left(\frac{s}{\rho}\right)
e^{i\tau\lambda(1-e^{-\frac{s}{\rho\lambda}})}\bigg\},
$$
take $1/\rho$ as the asymptotic parameter, and switch the roles of
$f$ and $h$ \cite{Handels} [$f(s)=J_0(s)$,
$h(s)=J_0(s)e^{i\tau\lambda(1-e^{-s/\lambda})}$]. The Mellin
transform of $h$ is given now by
$$
M[h;z]=\int_0^{\infty}
dt\,t^{z-1}J_0(t)e^{i\tau\lambda(1-e^{-t/\lambda})},
$$
which converges in the strip $0<\Re(z)<3/2$. We do not know any
closed expression for this integral but it suffices to obtain the
first term of the asymptotic expansion of the commutator when
$\rho\rightarrow0^+$. The asymptotic expansion of $f$ gives now
$a_k=2k$, ($k=0,\,1,\ldots$), $p_{00}=1,\,p_{10}=-1/4$,
\emph{etc.} and using the theorem we obtain the expansion
(\ref{r-int-0}). Under mild restrictions \cite{Handels} it can be
shown that $M[h;z]$ admits an analytic continuation for
$\Re(z)\rightarrow\infty$ which is meromorphic, at worst, in the
complex plane, and that can be used to get further terms of the
analytic expansion as $\rho\rightarrow0^+$. Notice, however, that
beyond the first term we cannot use the integral representation of
the Mellin transform of $h$ given above because the integral
becomes divergent for $z=2k$; $k\geq1$. This prevents us from
getting further terms in the asymptotic expansion as
$\rho\rightarrow0^+$ by using this method.

\subsection{\label{tau-math} Asymptotic behavior in $\tau$}

The integral in (\ref{newintegral})  is not an $h$-transform when
$\tau$ is chosen as the asymptotic parameter but can be written as
such by using the change of variables $u=\lambda e^{-t/\lambda}$.
This gives
\begin{equation}
\frac{\lambda }{R_1}\Im\bigg\{e^{i\tau\lambda}\int_0^{\lambda}du
\frac{e^{-i\tau u}}{u}
J_0\left(\lambda\log\frac{u}{\lambda}\right)
J_0\left(\rho\lambda\log\frac{u}{\lambda}\right)\bigg\}.\label{g001}
\end{equation}
Even though this integral has now the appropriate form with
$f(u)=\frac{1}{u}\chi_{[0,\lambda]}(u)
J_0(\lambda\log\frac{u}{\lambda})J_0(\rho\lambda\log\frac{u}{\lambda})$
and $h(u)=e^{-iu}$, the asymptotic behavior of $f(u)$ as
$u\rightarrow 0^+$ is not of the type considered in the theorem.
This has some unpleasant consequences, the Mellin transform of $f$
cannot be analytically continued as a meromorphic function on
$\mathbb{C}$, and we cannot use the expressions given in the
theorem for the asymptotic behavior of $h$-transforms. This forces
us to follow a more complicated approach. We will consider the
cases $\rho=0$ and $\rho\neq0$ separately.

\subsubsection{\label{r=0} $\rho=0$}

\begin{figure}
\hspace{0cm}\includegraphics[width=16.5cm]{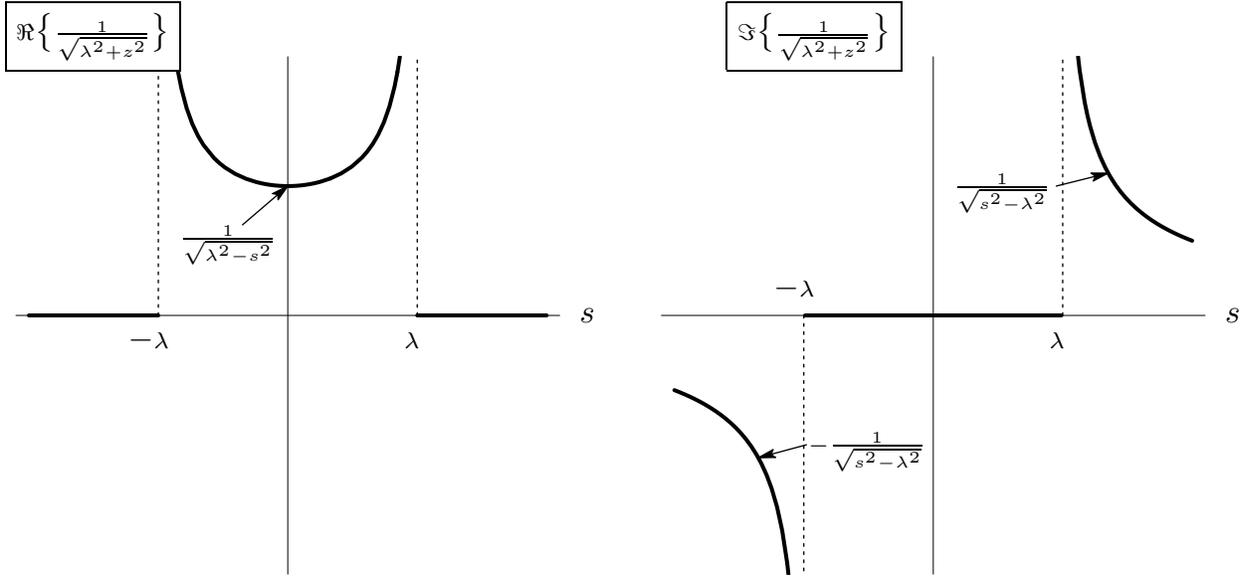}
\caption{Real and imaginary parts of
$\frac{1}{\sqrt{\lambda^2+z^2}}$ on the curve $z=-0^++is$
(corresponding to $\rho=0$). In the vicinity of the singularities
at $is=\pm i\lambda$ the real and imaginary parts of this function
behave as $\frac{1}{\sqrt{2\lambda(\lambda\mp
s)}}\quad(|s|<\lambda)$ and $\frac{\pm1}{\sqrt{2\lambda(\pm
s-\lambda)}}\quad(|s|>\lambda)$.} \label{fig4bis:cut}
\end{figure}

We have to study the integral
\begin{equation}
\frac{\lambda }{R_1}\Im\bigg\{e^{i\tau\lambda}\int_0^{\lambda}du
\frac{e^{-i\tau u}}{u}
J_0\left(\lambda\log\frac{u}{\lambda}\right)\bigg\}\label{tau1}
\end{equation}
when $\tau\rightarrow\infty$. Let us set $h(u)=e^{-i \tau u}$ and
$f(u)=\chi_{[0,\lambda]}(u)\frac{1}{u}J_0(\lambda\log\frac{u}{\lambda})$
[remember that $h$-transforms are defined as integrals over
$(0,\infty)$]. The Mellin transforms of these functions are
\begin{eqnarray*}
M[h;z]=\tau^{-z}\Gamma(z)e^{-i\pi z/2},
\end{eqnarray*}
convergent in the strip $0<\Re(z)<1$, and
\begin{eqnarray*}
M[f;z]=\frac{\lambda^{z-1}}{\sqrt{\lambda^2+(z-1)^2}},
\end{eqnarray*}
which converges in $1<\Re(z)$. Several comments are in order.
First we see that the Mellin transform of $f$ cannot be
analytically continued as a meromorphic function over the complex
plane because of the cuts coming from the square root. If we
choose the branch\footnote{We will use this choice throughout the
paper.} given by $\log(z)=\log|z|+i\arg(z)$ with $\arg(z)\in
(-\pi,\pi]$ the cuts are those parts of the imaginary axis with
$|z|\geq\lambda$. Another comment is that the regions where the
Mellin transforms converge do not overlap; this fact precludes us
from directly employing the Mellin-Parseval formula (\ref{th3}).
The reason can be traced back to the behavior of $h(t)$ as
$t\rightarrow 0^+$; fortunately this problem can be fixed in a
straightforward way. First we rewrite (\ref{tau1}) as
\begin{eqnarray*}
&&\frac{\lambda}{R_1}\Im\bigg\{e^{i\tau\lambda}\int_0^{\lambda}du
\frac{(e^{-i\tau u}-1)}{u}
J_0\left(\lambda\log\frac{u}{\lambda}\right)\bigg\}+
\frac{i\lambda}{R_1}\Im\bigg\{e^{i\tau\lambda}\int_0^{\lambda}du
\frac{1}{u}
J_0\left(\lambda\log\frac{u}{\lambda}\right)\bigg\}\\
&&=\frac{1}{R_1}\Im\bigg\{e^{i\tau\lambda}\Big[1+
\lambda\int_0^{\lambda}du \frac{(e^{-i\tau u}-1)}{u}
J_0\left(\lambda\log\frac{u}{\lambda}\right)\Big]\bigg\}.
\end{eqnarray*}
In order to study the last integral, we choose $f(u)$ as above and
$h(u)=\chi_{[0,\lambda]}(u)[e^{-i\tau u}-1]$. The Mellin transform
of $h$ can be written in terms of confluent hypergeometric
functions but one can work with a simpler expression by realizing
that, since $f(u)$ is zero if $u>\lambda$, one can replace $h$
with a new function $\tilde{h}$ that differs from it only for
$u>\lambda$:
\begin{eqnarray*}
\tilde{h}:\mathbb{R}^+\rightarrow\mathbb{R}:s\mapsto\left\{
\begin{array}{ll}
e^{-i\tau s}-1&s\in(0,\lambda)\\
e^{-i\tau s}&s\in[\lambda,\infty)
\end{array} \right. .
\end{eqnarray*}
The result is easily seen to be independent of this extension of
$h$. The integral giving the Mellin transform of $\tilde{h}$
converges in the strip $-1<\Re(z)<1$ with
$M[\tilde{h};z]=\tau^{-z}\Gamma(z)e^{-i\pi
z/2}-\frac{\lambda^z}{z}$. Since we now have overlapping
convergence strips  we can use the Mellin-Parseval formula and
obtain a Mellin-Barnes representation of our integral as
\begin{eqnarray}
&&\frac{1}{R_1}\Im\bigg\{e^{i\tau\lambda}\left[1+\frac{\lambda}{2\pi
i} \int_{c-i\infty}^{c+i\infty}\!\!dz
\frac{\tau^{-z}\lambda^{-z}\Gamma(z)e^{-i\pi z/2}-
\frac{1}{z}}{\sqrt{\lambda^2+z^2}}\right]\bigg\}\nonumber\\
&&=\frac{1}{R_1}\Im\bigg\{e^{i\tau\lambda}\left[1+\frac{\lambda}{2\pi
i} \int_{c-i\infty}^{c+i\infty}\!\!dz
\frac{\tau^{-z}\lambda^{-z}\Gamma(z)e^{-i\pi
z/2}}{\sqrt{\lambda^2+z^2}}\right]\bigg\} \label{mellinparseval}
\end{eqnarray}
with $\Re(c)\in(-1,0)$. We have employed the fact that
$$
\int_{c-i\infty}^{c+i\infty}\!\!dz
\frac{dz}{z\sqrt{\lambda^2+z^2}}=0.
$$
The analytic continuation of the integrand in the r.h.s. of
(\ref{mellinparseval}) is immediate. It has algebraic
singularities at $z=\pm i\lambda$ (see fig. \ref{fig4bis:cut})
with cuts in those points of the imaginary axis where
$|z|>\lambda$, and simple poles on $-z\in\mathbb{N}\cup\{0\}$.

In order to study the limit $\tau\rightarrow 0^+$ we can displace
the integration contour in (\ref{mellinparseval}) leftwards
parallel to the imaginary axis\footnote{Formula (\ref{th4}) is
obtained, precisely, by displacing the integration contour
parallel to the imaginary axis. For functions with the asymptotic
behaviors considered in the theorem the only singularities are
poles whose residues give the asymptotic expansion.}. A simple
analysis using the asymptotics of $\Gamma(z)$ for large values of
$\Im(z)$ shows that the series given by the residues of the
integrand at the poles $-z\in \mathbb{N}$, namely
\begin{eqnarray*}
\frac{\lambda}{R_1}\Im\bigg\{e^{i\tau\lambda}
\sum_{k=0}^{\infty}\frac{(-i\lambda\tau)^k}
{k!\sqrt{\lambda^2+k^2}}\bigg\},
\end{eqnarray*}
converges to the value of the integral. This is precisely what one
would get by expanding the exponential in (\ref{tau1}), exchanging
sum and integration, and computing the integrals that appear. This
was the procedure used in \cite{BarberoG.:2003ye}.

It is not possible to directly get the asymptotic behavior in the
$\tau\rightarrow\infty$ limit by displacing the integration
contour rightwards because of the presence of the cut and, even
worse, the absence of poles with $\Re(z)\geq0$. One could consider
choosing a value for $c$ and write (\ref{mellinparseval}) as a
real integral in the variable $y\in\mathbb{R}$ after the change
$z=c+iy$. In fact, by doing that one obtains the sum of two
$h$-transforms with asymptotic parameter $\log\tau$ that can be
studied with the standard formulas. Unfortunately, proceeding in
this way one gets a trivial \textit{zero} asymptotic expansion.
The reason, as we will find out later on, is that $\log\tau$ is
\textit{almost}, but not quite, the right asymptotic
parameter\footnote{More precisely, an asymptotic sequence given by
inverse powers of $\log\tau$ is appropriate to capture the
behavior of our integral in $\tau$.}.

\begin{figure}
\hspace{0cm}\includegraphics[width=11.5cm]{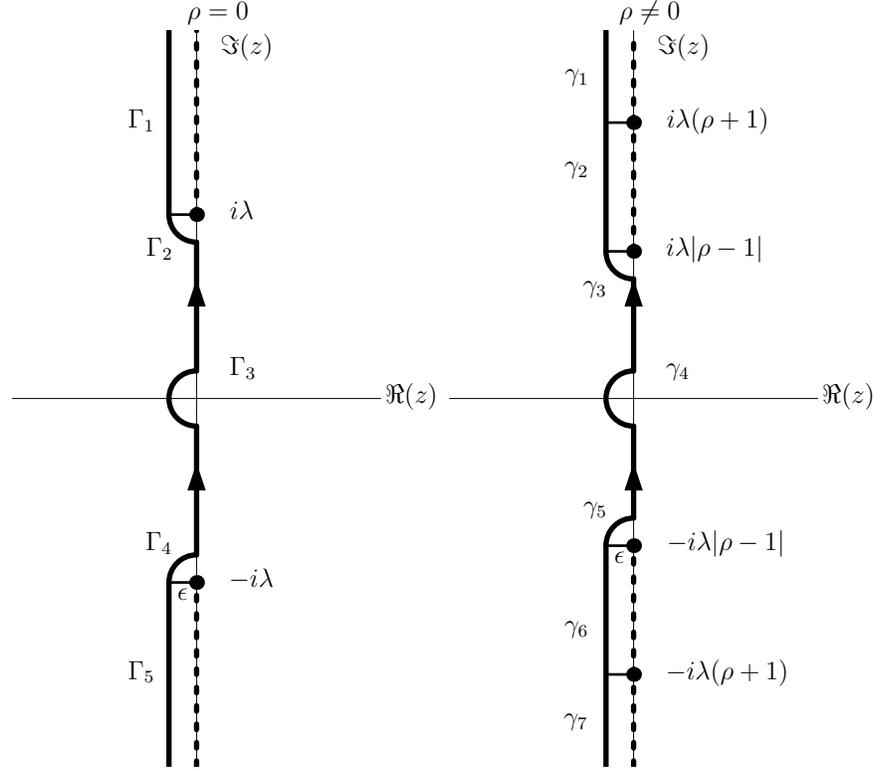}
\caption{Cuts and singularities of the integrand and integration
contours in the complex plane for the Mellin-Barnes representation
used in the asymptotic analysis for $\tau\rightarrow\infty$ in the
cases $\rho=0$ and $\rho\neq0$.} \label{fig5:cut}
\end{figure}

The solution to our problem is displacing the integration contour
all the way to the cut (choosing $c\rightarrow0^-$) as shown in
fig. \ref{fig5:cut}.
The contribution of the arcs $\Gamma_2$ and $\Gamma_4$ to the
integral in the r.h.s. of (\ref{mellinparseval}) goes to zero as
$-c=\epsilon\rightarrow0^+$. The remaining three contributions to
the integral (without prefactors) are
\begin{eqnarray*}
&&\Gamma_1\quad\quad-\int_{\lambda}^{\infty}dy
\frac{e^{-iy\log(\lambda\tau)}e^{\pi
y/2}\Gamma(iy)}{\sqrt{y^2-\lambda^2}},\\
&&\Gamma_3\quad\quad-\frac{\pi i}{\lambda}+i{\rm
P.V.}\int_{-\lambda}^{\lambda}dy
\frac{e^{-iy\log(\lambda\tau)}e^{\pi
y/2}\Gamma(iy)}{\sqrt{\lambda^2-y^2}},\\
&&\Gamma_5\quad\quad\int_{-\infty}^{-\lambda}dy
\frac{e^{-iy\log(\lambda\tau)}e^{\pi
y/2}\Gamma(iy)}{\sqrt{y^2-\lambda^2}}.
\end{eqnarray*}

The asymptotic expansions of these integrals can be obtained by a
variety of methods, for example by steepest descents. Nonetheless,
it is now possible to employ formula (\ref{th4}) and obtain, in
principle, the asymptotics to any order.

The integral over $\Gamma_1$ can be written as the following
standard $h$-transform by the change of variables $y=\lambda(1+s)$
with $s\in[0,\infty)$:
\begin{eqnarray*}
-e^{-i\lambda\log(\lambda\tau)}e^{\pi\lambda/2}\int_0^{\infty}ds\,e^{-i\lambda
s \log\tau}\frac{e^{-i\lambda s \log\lambda}e^{\pi\lambda
s/2}\Gamma[i\lambda(1+s)]}{\sqrt{s(s+2)}}.
\end{eqnarray*}
Taking $h(z)=e^{-i\lambda z}$, $f(z)=\frac{e^{-i\lambda z
\log\lambda}e^{\pi\lambda
z/2}\Gamma[i\lambda(1+z)]}{\sqrt{z(z+2)}}$, and the asymptotic
parameter $\log\tau$ we can use formulas (\ref{th3}) and
(\ref{th4}). Since the asymptotic behavior of $f$ when
$z\rightarrow 0^+$ is
$$
f(z)\sim \frac{\Gamma(i\lambda)}{\sqrt{2z}}+O(z^{1/2})
$$
and $M[h;z]=\lambda^{-z}e^{-\frac{\pi iz}{2}}\Gamma(z)$, we get
for the integral over $\Gamma_1$ an asymptotic expansion in powers
of $1/\sqrt{\log\tau}$ with a $\tau$ oscillating factor
$e^{-i\lambda\log(\lambda\tau)}$, namely
$$
-e^{-i\lambda\log(\lambda\tau)+\frac{\pi\lambda}{2}-\frac{\pi
i}{4}}\Gamma(i\lambda)\bigg\{\sqrt{\frac{\pi}{2\lambda\log\tau}}+
O\bigg[\frac{1}{(\log\tau)^{3/2}}\bigg]\bigg\}.
$$
A similar argument can be used to calculate the asymptotic
expansion of the integral over $\Gamma_5$,
$$
e^{i\lambda\log(\lambda\tau)-\frac{\pi\lambda}{2}+\frac{\pi
i}{4}}\Gamma(-i\lambda)\bigg\{\sqrt{\frac{\pi}{2\lambda\log\tau}}+
O\bigg[\frac{1}{(\log\tau)^{3/2}}\bigg]\bigg\}.
$$
Finally, to get the asymptotics of the integral over $\Gamma_3$ we
express it in the form
\begin{eqnarray*}
-\frac{\pi i}{\lambda}+i\int_0^{\lambda}dy
\cos[y\log\tau]\frac{e^{-i y\log\lambda}e^{\pi y/2}\Gamma(iy)+e^{i
y\log\lambda}e^{-\pi y/2}\Gamma(-iy)}{\sqrt{\lambda^2-y^2}}\hspace{.6cm}&&\\
-i\int_0^{\lambda}dy \sin[y\log\tau]\frac{e^{i
y\log\lambda}e^{-\pi y/2}\Gamma(-iy)-e^{-i y\log\lambda}e^{\pi
y/2}\Gamma(iy)}{\sqrt{\lambda^2-y^2}}\quad\,\,.&&
\end{eqnarray*}
By introducing neutralizers as in \cite{Handels} it is possible to
show that the critical points for the previous two integrals are 0
and $\lambda$. The contributions to the asymptotic expansion from
both points can be obtained in a straightforward way using formula
(\ref{th4}). The contribution from zero is $-\frac{\pi
i}{\lambda}$ whereas the one from $\lambda$ coincides precisely
with the sum of the contributions obtained before for $\Gamma_1$
and $\Gamma_5$. Summing all the terms and substituting the result
in (\ref{mellinparseval}) we finally get the behavior in $\tau$
presented in Section \ref{comm}.

It is worth remarking that the asymptotic expansions discussed can
be written as an oscillating factor that depends on $\log\tau$
multiplying an expansion in terms of inverse powers of
$\sqrt{\log\tau}$. Moreover, note that the frequency of these
oscillations is \emph{exactly} that of the oscillating factor that
we have already obtained.

\subsubsection{\label{rho dist 0} $\rho\neq0$}

\begin{figure}
\hspace{0cm}\includegraphics[width=16.5cm]{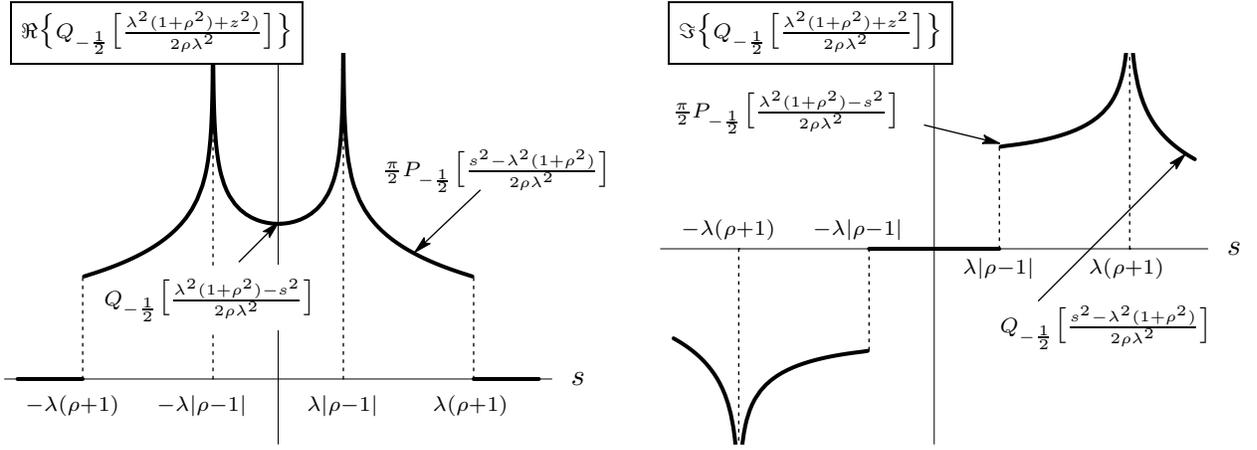}
\caption{Real and imaginary parts of
$Q_{-\frac{1}{2}}\big[\frac{\lambda^2(1+\rho^2)+z^2}{2\rho\lambda^2}\big]$
on the curve $z=-0^++is$ for $\rho\neq0$.} \label{fig5bis:cut}
\end{figure}

Consider now the integral (\ref{g001}). We set $h(u)=e^{-i \tau
u}$ and
$f(u)=\chi_{[0,\lambda]}(u)\frac{1}{u}J_0(\lambda\log\frac{u}{\lambda})
J_0(\rho\lambda\log\frac{u}{\lambda})$. The Mellin transform of
$f$ is
$$
M[f;z]=\frac{\lambda^{z-2}}{\pi\sqrt{\rho}}
Q_{-\frac{1}{2}}\left[\frac{\lambda^2(1+\rho^2)+(z-1)^2}{2\rho\lambda^2}\right].
$$
Proceeding as in the $\rho=0$ case we get the following
Mellin-Barnes representation:
\begin{eqnarray}
&&\frac{1}{\pi
R_1\sqrt{\rho}}\Im\bigg\{e^{i\tau\lambda}\bigg[Q_{-\frac{1}{2}}\left(\frac{1+\rho^2}
{2\rho}\right)+\frac{1}{2\pi i}
\int_{c-i\infty}^{c+i\infty}\!\!\!\!dz
\left(\tau^{-z}\lambda^{-z}\Gamma(z)e^{-i\pi z/2}-
\frac{1}{z}\right)Q_{-\frac{1}{2}}\left(\frac{\lambda^2(1+\rho^2)+z^2}
{2\rho\lambda^2}\right)\bigg]\bigg\}\nonumber\\
&&=\frac{1}{\pi
R_1\sqrt{\rho}}\Im\bigg\{e^{i\tau\lambda}\bigg[Q_{-\frac{1}{2}}\left(\frac{1+\rho^2}
{2\rho}\right)+\frac{1}{2\pi i}
\int_{c-i\infty}^{c+i\infty}\!\!\!\!dz\,\,
\tau^{-z}\lambda^{-z}e^{-i\pi
z/2}\Gamma(z)Q_{-\frac{1}{2}}\left(\frac{\lambda^2(1+\rho^2)+z^2}
{2\rho\lambda^2}\right)\bigg]\bigg\}.\label{mellinparsevalrho}
\end{eqnarray}
The main difference with the previous case is the appearance of
the associated Legendre function $Q_{-\frac{1}{2}}$. This function
has cuts on the imaginary axis for $|\Im(z)|\geq \lambda|\rho-1|$.
On the left of the cut the real and imaginary parts behave as
shown in fig. \ref{fig5bis:cut}. They become singular in the
neighborhood of the four points $z=\pm i\lambda|\rho-1|$ and
$z=\pm i\lambda(\rho+1)$. The real functions that describe each of
the continuous pieces are shown in the plots. Some of them are
given by $Q_{-\frac{1}{2}}(x)$ evaluated on $x>1$. Quite
surprisingly, the others can be written in terms of the associated
Legendre function $P_{-\frac{1}{2}}$
[$P_{-\frac{1}{2}}(x)=F\left(\frac{1}{2},\frac{1}{2};1;\frac{1-x}{2}\right)$]
with argument $x>-1$. The fundamental consequence of this is a
change in the singularity structure of the integrand of
(\ref{mellinparsevalrho}).
The limit $\tau\rightarrow0^+$ can be obtained as we did for
$\rho=0$. The result is given by (\ref{series}).

By displacing the integration contour rightwards to the cut as in
the $\rho=0$ case (see fig. \ref{fig5:cut}) we can split the last
integral in (\ref{mellinparsevalrho}) into seven different
contributions corresponding to the integration curves
$\gamma_1,\ldots,\gamma_7$. It is straightforward to show that the
contributions from $\gamma_3$ and $\gamma_5$ go to zero as the
contour gets closer to the imaginary axis. We are then left with
\begin{eqnarray*}
&&\gamma_1\quad\quad\hspace{1mm}
-\int_{\lambda(\rho+1)}^{\infty}\hspace{-6mm}dy\hspace{2mm}
e^{-iy\log(\lambda\tau)}e^{\pi
y/2}\Gamma(iy)Q_{-\frac{1}{2}}\left(\frac{y^2-\lambda^2(1+\rho^2)}
{2\rho\lambda^2}\right),\\
&&\gamma_2\quad\quad\frac{\pi}{2}i\int_{\lambda|\rho-1|}^{\lambda(\rho+1)}
\hspace{-6mm}dy\hspace{2mm} e^{-iy\log(\lambda\tau)}e^{\pi
y/2}\Gamma(iy)\bigg[
P_{-\frac{1}{2}}\left(\frac{y^2-\lambda^2(1+\rho^2)}
{2\rho\lambda^2}\right)+iP_{-\frac{1}{2}}\left(\frac{\lambda^2(1+\rho^2)-y^2}
{2\rho\lambda^2}\right)\bigg],\\
&&\gamma_4\quad\quad -i\pi Q_{-\frac{1}{2}}\left(\frac{1+\rho^2}
{2\rho}\right)+iP.V.\int_{-\lambda|\rho-1|}^{\lambda|\rho-1|}\hspace{-6mm}dy\hspace{2mm}
e^{-iy\log(\lambda\tau)}e^{\pi
y/2}\Gamma(iy)Q_{-\frac{1}{2}}\left(\frac{\lambda^2(1+\rho^2)-y^2}
{2\rho\lambda^2}\right),\\
&&\gamma_6\quad\quad\frac{\pi}{2}i\int_{-\lambda(\rho+1)}^{-\lambda|\rho-1|}
\hspace{-6mm}dy\hspace{2mm} e^{-iy\log(\lambda\tau)}e^{\pi
y/2}\Gamma(iy)\bigg[
P_{-\frac{1}{2}}\left(\frac{y^2-\lambda^2(1+\rho^2)}
{2\rho\lambda^2}\right)-iP_{-\frac{1}{2}}\left(\frac{\lambda^2(1+\rho^2)-y^2}
{2\rho\lambda^2}\right)\bigg],\\
&&\gamma_7\quad\quad\quad\int_{-\infty}^{-\lambda(\rho+1)}\hspace{-6mm}dy\hspace{2mm}
e^{-iy\log(\lambda\tau)}e^{\pi
y/2}\Gamma(iy)Q_{-\frac{1}{2}}\left(\frac{y^2-\lambda^2(1+\rho^2)}
{2\rho\lambda^2}\right).
\end{eqnarray*}
All these integrals can be written as Fourier transforms whose
only critical points are the integration limits. These can be
isolated by using the appropriate neutralizers \cite{Handels} and
the integrals so obtained can be written as $h$-transforms by
straightforward changes of variables. An interesting feature of
some of the integrals that arise is the fact that the asymptotic
expansion of the corresponding $f$ functions involve logarithms
precisely in the form assumed in (\ref{th2}). This means that in
addition to the $p_{m0}$ coefficients that appeared in the
$\rho=0$ case we will have also contributions coming from the
$p_{mn}$'s with $n\neq0$.

In order to get the asymptotic behavior of the different $f$
functions close to their singularities it is useful to remember
the asymptotic expansions of $P_{-\frac{1}{2}}(x)$ and
$Q_{-\frac{1}{2}}(x)$ at $x=-1^+$ and $x=1^+$, respectively:
$$
P_{-\frac{1}{2}}(x)\sim
\frac{5}{\pi}\log2-\frac{1}{\pi}\log(x+1)+\cdots
$$
$$
Q_{-\frac{1}{2}}(x)\sim
\frac{5}{2}\log2-\frac{1}{2}\log(x-1)+\cdots
$$

With the previous guidelines, and following the same steps as in
the $\rho=0$ case, it is easy (albeit lengthy) to deduce
(\ref{tauasimprneq0}). It is possible to arrive at the same result
by different methods that are somewhat simpler (i.e. steepest
descents) but they only allow to compute the first contribution to
the asymptotic behavior of the field commutator. The advantage of
the method employed here is that one can calculate as many terms
as desired for the asymptotic expansion just by taking more terms
in the asymptotic series of the function $f$.

\subsection{\label{lambda-math} Asymptotic behavior in $\lambda$}

We will consider now integrals of the form
\begin{equation}
\int_B^Adq\oint_{\gamma_1}dt_1 \oint_{\gamma_2}dt_2
f_{(0)}(q;t_1,t_2)e^{\lambda\Phi(q;t_1,t_2)} \label{Greenth}
\end{equation}
($0\leq B<A$) where $q$ is a real variable, $t_1$ and $t_2$ are
complex variables and $\gamma_{1,2}$ are closed paths (possibly
different) in the complex plane. The functions $f_{(0)}$ and
$\Phi$ are supposed to be holomorphic in the two complex arguments
and $C^n$ in $q$ (with $n$ possibly infinite) in an open
neighborhood of the integration region. In the following we will
use the notation $\partial_i\equiv\partial_{t_i},\,\,i=1,\,2$,
$\nabla\equiv(\partial_q,\partial_1,\partial_2)$, and
$||\nabla\Phi||^2\equiv
\eta_q(\partial_q\Phi)^2+\eta_{t_1}(\partial_1\Phi)^2+\eta_{t_2}(\partial_2\Phi)^2$,
where $\eta_q$, $\eta_{t_1}$, and $\eta_{t_2}$ are some functions
of $q$, $t_1$, and $t_2$ that will be chosen later in the
discussion. This freedom will prove to be useful when we
parametrize the curves $\gamma_{1,2}$ because it will allow us to
have denominators in the integrand that are real positive
functions, facilitating the analysis of the $A\rightarrow\infty$
limit. If $\|\nabla\Phi\|^2$ differs from zero for all
$q\in[B,A]$, $t_1\in\gamma_1$, and $t_2\in\gamma_2$, it is
possible to devise an ``integration by parts" procedure for this
type of integrals by writing them as
\begin{eqnarray*}
\frac{1}{\lambda}\int_B^Adq\oint_{\gamma_1}dt_1
\oint_{\gamma_2}dt_2\bigg[\frac{f_{(0)}\eta_q\partial_q\Phi
\partial_qe^{\lambda\Phi}}{\|\nabla\Phi\|^2}+\frac{f_{(0)}\eta_{t_1}\partial_1\Phi
\partial_1e^{\lambda\Phi}}{\|\nabla\Phi\|^2}+\frac{f_{(0)}\eta_{t_2}\partial_2\Phi
\partial_2e^{\lambda\Phi}}{\|\nabla\Phi\|^2}\bigg].
\end{eqnarray*}
The contribution of the first term can be written, after
integrating by parts in $q$, as
\begin{eqnarray*}
\frac{1}{\lambda}\oint_{\gamma_1}dt_1 \oint_{\gamma_2}dt_2
\frac{f_{(0)}\eta_q\partial_q\Phi
e^{\lambda\Phi}}{\|\nabla\Phi\|^2}\Bigg|_B^A-
\frac{1}{\lambda}\oint_{\gamma_1}dt_1
\oint_{\gamma_2}dt_2\int_B^Adq
e^{\lambda\Phi}\partial_q\frac{f_{(0)}\eta_q\partial_q\Phi}{\|\nabla\Phi\|^2}
\end{eqnarray*}
(notice that the integration region is compact and the function
integrable so that we can apply Fubini's theorem and freely
exchange orders of integration). Integration by parts in $t_1$
gives
\begin{eqnarray*}
\frac{1}{\lambda}\int_B^Adq\oint_{\gamma_2}dt_2
\oint_{\gamma_1}dt_1
\partial_1\frac{f_{(0)}\eta_{t_1}\partial_1\Phi e^{\lambda\Phi}}{\|\nabla\Phi\|^2}-
\frac{1}{\lambda}\int_B^Adq\oint_{\gamma_2}dt_2
\oint_{\gamma_1}dt_1
e^{\lambda\Phi}\partial_1\frac{f_{(0)}\eta_{t_1}\partial_1\Phi}{\|\nabla\Phi\|^2}.
\end{eqnarray*}
If $(f_{(0)}\eta_{t_1}\partial_1\Phi
e^{\lambda\Phi})\|\nabla\Phi\|^{-2}$ is a holomorphic function of
$t_1$ in an open set containing the \emph{closed} curve $\gamma_1$
(for all $q\in[B,A]$ and $t_2\in\gamma_2$) the first integral is
zero and we are left only with the second term. The same argument
applies to the integration in $t_2$. We finally get that
(\ref{Greenth}) can be written as
\begin{equation}
\frac{1}{\lambda}\oint_{\gamma_1}dt_1 \oint_{\gamma_2}dt_2
\frac{f_{(0)}\eta_q
\partial_q\Phi e^{\lambda\Phi}}{\|\nabla\Phi\|^2}\Bigg|_B^A-
\frac{1}{\lambda}\int_B^Adq\oint_{\gamma_1}dt_1
\oint_{\gamma_2}dt_2 f_{(1)}(q;t_1,t_2)e^{\lambda\Phi(q;t_1,t_2)},
\label{intbyparts}
\end{equation}
here we have introduced the notation
$$
f_{(k+1)}\equiv\partial_q\frac{f_{(k)}\eta_q\partial_q\Phi}{\|\nabla\Phi\|^2}
+\partial_1\frac{f_{(k)}\eta_{t_1}\partial_1\Phi}{\|\nabla\Phi\|^2}+
\partial_2\frac{f_{(k)}\eta_{t_2}\partial_2\Phi}{\|\nabla\Phi\|^2}\quad k=0,\,1,\ldots.
$$
The second integral in (\ref{intbyparts}) has the same structure
as the original integral, so the same procedure can be applied as
long as the appropriate regularity conditions are satisfied for
expressions involving the successive $f_{(k)}$, $k=1,\ldots$ The
strategy to obtain asymptotic expansions for the original integral
consists in getting expansions for the lower dimensional boundary
integrals that arise in this way, provided that the remaining
non-boundary integral behaves nicely in the
$\lambda\rightarrow\infty$ limit. Particularizing this argument to
the case of a single complex variable $t$ is straightforward.

In our case [see equation (\ref{trintegral})] we actually want to
obtain an asymptotic expansion of an improper integral. So we will
also have to study the asymptotics of the limit
$A\rightarrow\infty$,
\begin{eqnarray}
&&-\frac{1}{\lambda}\oint_{\gamma_1}dt_1 \oint_{\gamma_2}dt_2
\frac{f_{(0)}\eta_q
\partial_q\Phi e^{\lambda\Phi}}{\|\nabla\Phi\|^2}\Bigg|_B\label{intbypartsimp}\\
&&+\lim_{A\rightarrow\infty}\bigg\{\frac{1}{\lambda}\oint_{\gamma_1}dt_1
\oint_{\gamma_2}dt_2 \frac{f_{(0)}\eta_q
\partial_q\Phi e^{\lambda\Phi}}{\|\nabla\Phi\|^2}\Bigg|_A-
\frac{1}{\lambda}\int_B^Adq\oint_{\gamma_1}dt_1
\oint_{\gamma_2}dt_2
f_{(1)}(q;t_1,t_2)e^{\lambda\Phi(q;t_1,t_2)}\bigg\}.\nonumber
\end{eqnarray}

In order to derive useful asymptotic expansions from the previous
formulas it is desirable that $e^{\lambda\Phi(q;t_1,t_2)}$ be
bounded by a $\lambda$-independent function\footnote{The
asymptotic analysis of multiple integrals is greatly simplified by
the identification of the \emph{critical points}, those points
that give the dominant contributions. For Laplace or Fourier types
of integrals these are easy to identify and, in practice, they are
just a finite number of isolated points that can be singled out
and studied by using neutralizers.}. This happens for example if
$\Re[\Phi(q;t_1,t_2)]\leq0$ in the integration region
$[B,A]\times\gamma_1\times\gamma_2$. In fact, we will take
advantage of the freedom in the choice of the integration contours
$\gamma_{1,2}$ to impose this condition.

If there are points in the integration region (referred to in the
following as singular points) where $\|\nabla\Phi\|^2=0$ the
argument presented above to derive (\ref{intbyparts}) is no longer
true. In such a situation the way to proceed \cite{Handels,
Vander} is to isolate them by introducing
neutralizers\footnote{These are real, positive,
$C^{\infty}(\mathbb{R}$) functions
$\nu(q;\alpha_1,\alpha_2;\beta_1,\beta_2)$ satisfying
$\nu(q;\alpha_1,\alpha_2;\beta_1,\beta_2)=0$ if
$q\in(-\infty,\alpha_1]\cup[\alpha_2,\infty)$ and
$\nu(q;\alpha_1,\alpha_2;\beta_1,\beta_2)=1$ if
$q\in[\beta_1,\beta_2]$ (some of the parameters
$\alpha_1,\alpha_2,\beta_1,\beta_2$ may be taken to be infinite).
In several variables $q_1,\ldots q_n$ it is sometimes useful to
consider neutralizers that depend only on $\|q\|$.}. This allows
us to divide the integration region into several pieces and study
the resulting integrals by choosing the most suitable techniques.
This is facilitated by the fact that neutralizers force most of
the boundary integrals to be zero. In the process of obtaining
asymptotic expansions from (\ref{intbypartsimp}) it may be useful
to further divide the integration region of the last integral in
several pieces, by introducing additional neutralizers, in order
to localize the singular points in a region that does not contain
the boundary $q=A$; this simplifies the analysis of the
$A\rightarrow\infty$ limit.

The freedom to choose the integration contours in (\ref{Greenth})
may be useful for other purposes. One can, for example, avoid
singularities of the boundary integrals over
$\{B\}\times\gamma_1\times\gamma_2$ and
$\{A\}\times\gamma_1\times\gamma_2$; it may be possible to write
the integrals involving the singular points as (multidimensional)
Laplace or Fourier transforms for which the analysis of singular
points can be carried out by standard procedures (see
\cite{Handels}) or to improve the behavior for large values of $q$
if one intends to study the asymptotics of the improper integral
when $A\rightarrow\infty$\footnote{It is important to point out
that the asymptotic behavior of an improper integral over
$[0,\infty)$ may be very different from the limit of the
asymptotic expansion of the integral over $[0,A]$ when
$A\rightarrow\infty$.}. Finally in some cases this freedom may
allow us to simplify the analysis of the contribution of some of
the singular points.

As in the study of the $\tau$ asymptotics, we will consider the
cases $\rho=0$ and $\rho\neq0$ separately.

\subsubsection{\label{rholamb 0} $\rho=0$}

\begin{figure}
\hspace{0cm}\includegraphics[width=11.5cm]{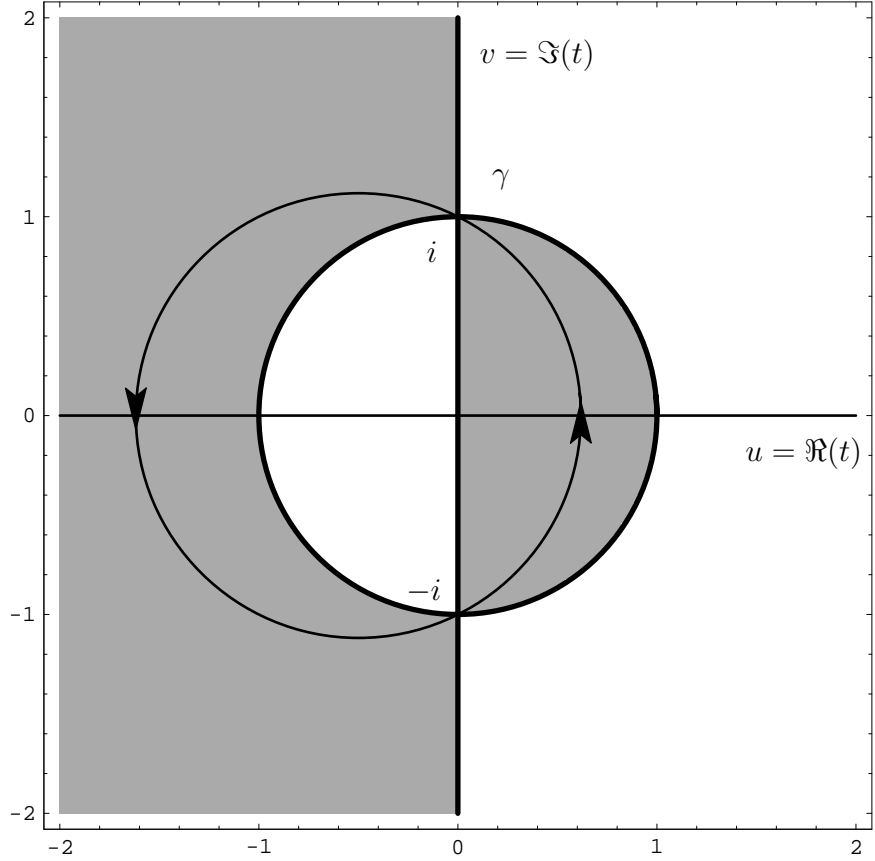}
\caption{The region $\Re(t-1/t)\leq0$ consists of the shadowed
area and its boundary. A possible choice of the integration
contour $\gamma$ is also shown.} \label{fig7:gamma}
\end{figure}

We study the integral (\ref{trintegral}) for $\rho=0$:
\begin{equation}
-\frac{1}{R_1}\lim_{A\rightarrow\infty}\Im\bigg\{\frac{i\lambda
e^{i\tau\lambda}}{2\pi}\int_0^{A}dq\oint_{\gamma} dt\frac{1}{t}
e^{\lambda\big[\frac{\scriptstyle
q}{\scriptstyle2}(t-\frac{\scriptstyle1}{\scriptstyle t})-i\tau
e^{-q}\big]}\bigg\}.\label{bintegral}
\end{equation}
Let us discuss first the possible choices of integration contour
$\gamma$. As commented above it is convenient to impose that the
real part of the exponent in the integrand be less or equal to
zero, i.e. $\Re(t-1/t)\leq0$. It is straightforward to show that
this happens when $u(u^2+v^2-1)\geq0$ ($t=u+iv$). This complex
region (see fig. \ref{fig7:gamma}) is given by the points inside
the unit circle centered at the origin with positive real part,
those outside this circle with negative real part, and the
boundary of these two regions (i.e. the imaginary axis and the
unit circumference centered at the origin). The integration
contour can be any closed, positively oriented, simple curve
contained in this region that surrounds the origin. Notice that
all these contours contain the points $t=\pm i$.
If we choose $\eta_q=-1$ and $\eta_t=t^2$ we have
$$\|\nabla\Phi\|^2=\frac{1}{4}\left[q^2\,{\left( t +
\frac{1}{t} \right) }^2 - {\left( t-\frac{1}{t} + 2i\tau e^{-q}
\right)}^2\right].$$ At the left boundary ($q=0$) this is zero
when $t=-i\tau\pm\sqrt{1-\tau^2}$. Depending on the value of
$\tau$ these points may be on the unit circumference; it is,
hence, convenient to choose integration contours in the allowed
part of $\mathbb{C}$ that avoid them (see fig. \ref{fig7:gamma}).
This will always be possible except for $\tau=1$. When $q\neq0$
the condition $\|\nabla\Phi\|^2=0$ gives, generically, four
possible solutions (depending on $q$ and $\tau$) that may or may
not be in the integration region. If $\tau>1$, there is one of
these, ($t=-i$, $q=\log\tau$) that will be in the integration
region for every possible choice of the integration contour
$\gamma$. The remaining ones will or will not belong to the
integration region depending on $\gamma$. In principle one would
proceed by picking a certain contour, determine the values of $q$
corresponding to these points, introduce suitable neutralizers,
and study the neutralized integrals one at a time. There is,
however, a simple argument that shows that it is not necessary to
discuss them in detail (we assume $q\neq0$ in the following
discussion unless otherwise specified).

The idea is to write the integral in (\ref{bintegral}) as the sum
of three integrals $I_j$, $j=1,\,2,\,3$ and choose
\emph{different}\footnote{The reason why we use neutralizers
depending only on $q$ is, precisely, to exploit this freedom.}
contours for each of them. These integrals are
\begin{eqnarray*}
&&I_j\equiv-\frac{1}{R_1}\Im\Big\{\frac{i\lambda
e^{i\tau\lambda}}{2\pi}\int_0^{A}dq\oint_{\gamma}
dt\,\nu_j(q)\frac{1}{t} e^{\lambda\big[\frac{\scriptstyle
q}{\scriptstyle2}(t-\frac{\scriptstyle1}{\scriptstyle t})-i\tau
e^{-q}\big]}\Big\}
\end{eqnarray*}
where we have introduced three  neutralizer functions $\nu_j(q)$,
$j=1,\,2,\,3$ satisfying $\nu_1+\nu_2+\nu_3=1$ in $[0,A]$ and
\begin{eqnarray*}
&&\nu_1(q)=1\quad \textrm{if}\quad q\in[0,\alpha_1],\\
&&\nu_1(q)=0\quad \textrm{if}\quad q\in[\alpha_2,A],\\
&&\nu_2(q)=0\quad \textrm{if}\quad q\in[0,\alpha_1]\cup[\beta_2,A],\\
&&\nu_2(q)=1\quad \textrm{if}\quad q\in[\alpha_2,\beta_1],\\
&&\nu_3(q)=0\quad \textrm{if}\quad q\in[0,\beta_1],\\
&&\nu_3(q)=1\quad \textrm{if}\quad q\in[\beta_2,A],
\end{eqnarray*}
with $0<\alpha_1<\alpha_2<\beta_1<\beta_2<A$ (the specific choices
for these parameters will be explained later). By doing this the
effective integration regions in $q$ are $[0,\alpha_2]$,
$[\alpha_1,\beta_2]$ and $[\beta_1,A]$ respectively; notice that
the boundary $q=0$ appears only in the first.

Let us consider first $I_1$. In this case it is convenient to
choose an integration contour $\gamma$ that avoids the unit
circumference (except at $t=\pm i$) because, if $\tau<1$, the
integration by parts procedure introduced above could, otherwise,
give a boundary contribution at $q=0$ involving a singular
function. We pick the contour depicted in fig. \ref{fig7:gamma}.
In order to use integration by parts it is also necessary to show
that it is possible to find $\alpha_2$ such that
$\|\nabla\Phi\|^2\neq0$ in the integration region. The solutions
to $\|\nabla\Phi\|^2=0$ are those satisfying the following
quadratic equations
\begin{eqnarray*}
&& (1+q)t^2+2i\tau e^{-q}t-1+q=0, \\
&& (1-q)t^2+2i\tau e^{-q}t-1-q=0 .
\end{eqnarray*}
It can easily be shown (by using the implicit function theorem or
the explicit solution of these equations in $t$) that the
solutions are continuous at $q=0$ (if $\tau\neq1$), i.e. the
solutions for small enough $q$ are close to
$t=-i\tau\pm\sqrt{1-\tau^2}$. If $\tau\neq1$ it is then possible
to choose $\alpha_2$ and $\gamma$ in such a way that
$\|\nabla\Phi\|^2\neq0$ in the integration region
$\gamma\times[0,\alpha_2]$.

We can use formula (\ref{intbypartsimp}) to get an asymptotic
expansion for $I_1$. We first note that the term corresponding to
the second integral in (\ref{intbyparts}) is $O(1/\lambda)$
because our choice of contour allows us to bound the absolute
value of the integrand by a regular function independent of
$\lambda$. The integral $I_1$ is then given at leading order by
the contribution of the boundary terms which can be written as
$$
I_1=\frac{1}{\pi R_1}\Im\bigg\{i\oint_{\gamma}\frac{dt}{t^2+2i\tau
t-1} \bigg\}
$$
[the function $\nu_1(q)$ kills the contribution coming from the
$q=\alpha_2$ boundary]. If $\tau<1$ the poles of the integrand
$t=-i\tau\pm\sqrt{1-\tau^2}$ are on the unit circumference but
only that with the negative real part is inside $\gamma$. The
value of the residue of the integrand at this point is
$-\frac{1}{2\sqrt{1-\tau^2}}$ and hence
$$
I_1=\frac{1}{R_1}\Im\bigg\{\frac{1}{\sqrt{1-\tau^2}}\bigg\}=0.
$$
If $\tau>1$ the poles of the integrand are on the imaginary axis
$t=-i\tau\pm i\sqrt{\tau^2-1}$; now only that with the positive
sign in the square root is inside $\gamma$. The residue is
$\frac{1}{2i\sqrt{\tau^2-1}}$ and hence
\begin{equation}
I_1=\frac{1}{R_1}\Im\bigg\{\frac{i}{\sqrt{\tau^2-1}}\bigg\}
=\frac{1}{R_1\sqrt{\tau^2-1}}.\label{ff1}
\end{equation}
Consider next $I_2$. In this case it is convenient to choose the
unit circumference for $\gamma$ because, parameterizing it by an
angle ($t(\theta)=e^{i\theta}$, $\theta\in(-\pi,\pi]$), the
exponent $\lambda\Phi(q;t)$ becomes a purely imaginary function of
the two real variables $q$ and $\theta$,
$$
i\lambda\phi(q;\theta)\equiv\lambda\Phi[q;t(\theta)]=i\lambda(q\sin\theta-\tau
e^{-q}),
$$
and the integral becomes a Fourier transform on a rectangular
region determined by the support of the neutralizer. As is well
known the critical points for Fourier transforms are given by the
boundaries of the integration region and those interior points
where the gradient of the \emph{real} function $\phi$ is zero. In
our case  the $q=\alpha_1$ and $q=\beta_2$ boundaries give no
contribution for this integral owing to the presence of the
neutralizer. Let us then find the interior critical points. Since
$\|\nabla\Phi\|^2[q;t(\theta)]$ is a sum of squares of real
numbers\footnote{This is no longer true for other choices of
$\gamma$ for which $\|\nabla\Phi\|^2$ is, in general, complex. In
this cases $\|\nabla\Phi\|^2$ may be zero even though the partial
derivatives of $\Phi$ are different from zero.} these points are
precisely those where our procedure of integration by parts fails.
We have to solve the following equations
\begin{eqnarray*}
&&q\cos\theta=0,\\
&&\sin\theta+\tau e^{-q}=0.
\end{eqnarray*}
The solution is $q=0$, $\theta=-\arcsin\tau$ if $\tau<1$ (these
are outside the effective integration region for $I_2$) and
$q=\log\tau$, $\theta=-\frac{\pi}{2}$ if $\tau>1$. The choices of
$\alpha_{1,2}$ can be made in such a way that the critical point
with $q>0$ (present only when $\tau>1$) always corresponds to
$I_2$  (we also choose $\beta_{1}$ such that $\log\tau<\beta_1$ to
avoid having critical points in $I_3$). The asymptotic expansion
of $I_2$ can be now obtained by using the standard formulas
\cite{Handels} for the contribution of critical points. In our
case this is simply
\begin{equation}
\frac{1}{R_1}\Im\bigg\{
\frac{e^{i\lambda(\tau-\log\tau-1)}}{\sqrt{\log\tau}}\bigg\}+
O\bigg[\frac{1}{\lambda}\bigg]\hspace{1cm}\label{ff2}
\end{equation}
for $\tau>1$ and 0 if $\tau<1$. We do not need to discuss the
limit $A\rightarrow\infty$ of $I_1$ and $I_2$ because the
integrand has compact support independent of $A$; we have only
assumed that $\log\tau<A$. However this is no longer true for
$I_3$.

Let us analyze, finally, $\lim_{A\rightarrow\infty}I_3$. We choose
again the unit circumference for $\gamma$ (with the same
parametrization) and write
\begin{eqnarray*}
\lim_{A\rightarrow\infty}I_3=\Im\bigg\{\frac{\lambda
e^{i\tau\lambda}}{2\pi
R_1}\lim_{A\rightarrow\infty}\int_{\beta_1}^Adq\int_{-\pi}^{\pi}d\theta\,\nu_3(q)
e^{i\lambda\phi(q;\theta)}\bigg\}.
\end{eqnarray*}
Using formula (\ref{intbypartsimp}) with $\eta_q=-1$,
$\eta_t=t^2$, and taking into account that $\nu_3(\beta_1)=0$,
this expression becomes
\begin{equation}
\Im\bigg\{\frac{\lambda e^{i\tau\lambda}}{2\pi
R_1}\lim_{A\rightarrow\infty}\bigg[\frac{-i}{\lambda}\int_{-\pi}^{\pi}d\theta
\frac{(\tau
e^{-A}+\sin\theta)e^{i\lambda\phi(A;\theta)}}{A^2\cos^2\theta+(\tau
e^{-A}+\sin\theta)^2}+ \frac{i}{\lambda}
\int_{\beta_1}^{A}dq\int_{-\pi}^{\pi}d\theta\,\nu_3(q)
e^{i\lambda\phi(q;\theta)}g_1(q;\theta)\bigg]\bigg\}.\label{improper2}
\end{equation}
Here\footnote{In the following we will denote $g_k(q;\theta)\equiv
ie^{i\theta}f_{(k)}[q;t(\theta)]$, the $ie^{i\theta}$ factor comes
from the measure in the integral.}
$$
g_1(q;\theta)\equiv ie^{i\theta}f_{(1)}[q;t(\theta)]=\frac{(\tau
e^{-q}+\sin\theta)^2(\tau e^{-q}-
q\sin\theta)+q\cos^2\theta[(q^2-4)\sin\theta-\tau(q+4)e^{-q}]}
{[q^2\cos^2\theta+(\tau e^{-q}+\sin\theta)^2]^2}.
$$
We have disregarded in equation (\ref{improper2}) a term
containing derivatives of $\nu_3(q)$ because it can be shown to be
$O(1/\lambda^R)$ for every $R>0$. We want to show that
(\ref{improper2}) does not contribute to the asymptotic expansion
at the order in $\lambda$ considered. Note that it is difficult to
prove that the limit of the last integral is zero when
$\lambda\rightarrow\infty$  by using an argument inspired in the
Riemann-Lebesgue lemma, because changing variables to write it as
a Fourier transform introduces singularities in the integrand that
are not easy to deal with. To study the asymptotics of
(\ref{improper2}) we will follow instead several steps:

\noindent\textbf{i)} Use integration by parts to further decompose
the second integral in (\ref{improper2}) as a surface term and a
double integral in $q$ and $\theta$.

\noindent\textbf{ii)} Split each of the integrands obtained in
this way in two pieces; one with a simpler denominator and another
whose limit $A\rightarrow\infty$ vanishes. After this step we will
be left with two integrals in $q$ and a double integral.

\noindent\textbf{iii)} Split in two the double integral by writing
the exponential term as
$$
e^{i\lambda\phi(q;\theta)}=e^{i\lambda[q\sin\theta-\tau
e^{-q}]}=e^{i\lambda q\sin\theta}(e^{-i\lambda\tau
e^{-q}}-1)+e^{i\lambda q\sin\theta}.
$$

\noindent\textbf{iv)} Show that the contribution coming from
$e^{i\lambda q\sin\theta}(e^{-i\lambda\tau e^{-q}}-1)$ vanishes
sufficiently fast when $\lambda\rightarrow\infty$ by using the
H\"{o}lder inequality.

\noindent\textbf{v)} Finally, show that the remaining terms have a
simple asymptotic behavior.

\bigskip
\noindent\textbf{Step i)} After an additional integration by parts
(\ref{improper2}) becomes
\begin{eqnarray}
\Im\bigg\{\frac{\lambda e^{i\tau\lambda}}{2\pi
R_1}\lim_{A\rightarrow\infty}\bigg[-\frac{ie^{-i\lambda\tau
e^{-A}}}{\lambda}\int_{-\pi}^{\pi}d\theta \frac{(\tau
e^{-A}+\sin\theta)e^{i\lambda A\sin\theta}}{A^2\cos^2\theta+(\tau
e^{-A}+\sin\theta)^2}\hspace{1.3cm}&&\nonumber\\
+\frac{e^{-i\lambda\tau
e^{-A}}}{\lambda^2}\int_{-\pi}^{\pi}d\theta \frac{(\sin\theta-\tau
e^{-A})e^{i\lambda A\sin\theta}}{A^2\cos^2\theta+(\tau
e^{-A}+\sin\theta)^2}g_1(A;\theta)&&\nonumber\\
-\frac{1}{\lambda^2}
\int_{\beta_1}^{A}dq\int_{-\pi}^{\pi}d\theta\nu_3(q)
e^{i\lambda\phi(q;\theta)}g_2(q;\theta)\bigg]\bigg\}.\hspace{.7cm}&&\nonumber
\end{eqnarray}

\bigskip
\noindent\textbf{Step ii)} All the integrands that appear in the
first step have factors of the type
$$
\frac{1}{[q^2\cos^2\theta+(\tau e^{-q}+\sin\theta)^2]^N}\quad
N=1,\,2,\ldots
$$
that can be written as
$$
\frac{1}{[q^2\cos^2\theta+\sin^2\theta]^N}+\frac{\sum_{j=1}^{2N}\tau^j
e^{-jq}P_j(q,\sin\theta,\cos\theta)}{[q^2\cos^2\theta+(\tau
e^{-q}+\sin\theta)^2]^N [q^2\cos^2\theta+\sin^2\theta]^N}\quad
N=1,\,2,\ldots
$$
with $P_j(q,\sin\theta,\cos\theta)$ a polynomial of degrees
$2(N-1)$, $2(N-1)$, and  $2N-1$ in $q$, $\cos\theta$, and
$\sin\theta$ respectively. For fixed values of $q$ larger than a
certain $Q>\log\tau$ the maximum of the function
$y(\theta)=[q^2\cos^2\theta+(\tau e^{-q}+\sin\theta)^2]^{-1}$ is
$(1-\tau e^{-q})^{-2}$, so we have\footnote{We choose
$\beta_1>Q>\log\tau$ in the neutralizers introduced before.}
$$
\bigg|\frac{\sum_{j=1}^{2N}\tau^j
e^{-jq}P_j(q,\sin\theta,\cos\theta)}{[q^2\cos^2\theta+(\tau
e^{-q}+\sin\theta)^2]^N
[q^2\cos^2\theta+\sin^2\theta]^N}\bigg|\leq
\frac{\sum_{j=1}^{2N}\tau^je^{-jq} \tilde{P}_j(q)}{(1-\tau
e^{-q})^{2N}(q^2\cos^2\theta+\sin^2\theta)^N}\quad N=1,\,2,\ldots
$$
where $\tilde{P}_j(q)$ is the polynomial in $q$ obtained from
$P_j(q,\sin\theta,\cos\theta)$ by taking the absolute value of the
coefficients and substituting $\sin\theta=1$, $\cos\theta=1$. This
formula, and the fact that
$$
\int_{-\pi}^{\pi}d\theta\frac{1}{[A^2\cos^2\theta+\sin^2\theta]^N}\sim
2\frac{\sqrt{\pi}}{A}\frac{\Gamma(N-1/2)}{\Gamma(N)}\quad
\textrm{as}\,\,\, A\rightarrow\infty,
$$
allows us to show that the first two integrals obtained in step i)
can be substituted by
\begin{eqnarray*}
-\frac{i}{\lambda} \int_{-\pi}^{\pi}d\theta \frac{
\sin\theta}{A^2\cos^2\theta+\sin^2\theta}e^{i\lambda A\sin\theta}
+\frac{1}{\lambda^2}\int_{-\pi}^{\pi}d\theta
\frac{A(A^2-4)\sin^2\theta\cos^2\theta-A\sin^4\theta}
{[A^2\cos^2\theta+\sin^2\theta]^3}e^{i\lambda A\sin\theta}.
\end{eqnarray*}
Let us consider now the double integral appearing in step i). The
function $g_2(q;\theta)$ can be written as the sum of two pieces
$h_2(q;\theta)$ and $G_2(q;\theta)$. The first one is
\begin{eqnarray*}
&&h_2(q;\theta)=\\
&&\frac{q^4(q^2-4)\cos^6\theta+
q^2(40-23q^2+3q^4)\cos^4\theta\sin^2\theta-
(4-23q^2+8q^4)\cos^2\theta\sin^4\theta+(q^2-1)\sin^6\theta}
{(q^2\cos^2\theta+\sin^2\theta)^4}
\end{eqnarray*}
and $G_2(q;\theta)$  is a sum of factors $(\tau e^{-q})^N$
($N=1,\,2,\,\ldots$) times quotients of polynomials in $q$,
$\sin\theta$, and $\cos\theta$ divided by products of powers of
$[q^2\cos^2\theta+\sin^2\theta]$ and $[q^2\cos^2\theta+(\tau
e^{-q}\sin\theta)^2]$ (notice that these functions never vanish in
the integration region). The integral becomes
$$
-\frac{1}{\lambda^2}
\int_{\beta_1}^{A}dq\int_{-\pi}^{\pi}d\theta\,\nu_3(q)
e^{i\lambda\phi(q;\theta)}\big[h_2(q;\theta)+G_2(q;\theta)\big].
$$
The absolute value of the integrand involving $G_2(q;\theta)$ is
an integrable function owing to the decreasing exponential factors
and the fact that they multiply terms that grow polynomially at
worst in $q$. Hence its contribution to the total integral is
$O(1/\lambda^2)$ even after taking the limit $A\rightarrow\infty$.
We thus conclude that
\begin{eqnarray}
\Im\bigg\{\frac{\lambda e^{i\tau\lambda}}{2\pi
R_1}\lim_{A\rightarrow\infty}\bigg[-\frac{i}{\lambda}\int_{-\pi}^{\pi}d\theta
\frac{\sin\theta}{A^2\cos^2\theta+\sin^2\theta}e^{i\lambda
A\sin\theta}\hspace{2.9cm}&&\nonumber\\
+\frac{1}{\lambda^2} \int_{-\pi}^{\pi}d\theta
\frac{A(A^2-4)\sin^2\theta\cos^2\theta-A\sin^4\theta}
{[A^2\cos^2\theta+\sin^2\theta]^3}e^{i\lambda
A\sin\theta}\hspace{.5cm}
&&\label{improper4}\\
-\frac{1}{\lambda^2} \int_{\beta_1}^{A}dq\int_{-\pi}^{\pi}d\theta
e^{i\lambda\phi(q;\theta)}\nu_3(q)h_2(q;\theta)\bigg]\bigg\}+
O\bigg[\frac{1}{\lambda}\bigg].\hspace{.7cm}&&\nonumber
\end{eqnarray}

\bigskip
\noindent\textbf{Step iii)} Note that we have already managed to
substitute the exponential factor $e^{i\lambda\phi(A;\theta)}$ by
$e^{i\lambda\sin\theta}$ within the integrals over $\theta$.
However it is not possible to remove the function $\phi(q,\theta)$
from the double integral. We instead proceed to rewrite it as
$$
-\frac{1}{\lambda^2} \int_{\beta_1}^{A}dq\int_{-\pi}^{\pi}d\theta
e^{i\lambda q\sin\theta}[e^{-i\lambda\tau
e^{-q}}-1]\nu_3(q)h_2(q;\theta)-\frac{1}{\lambda^2}
\int_{\beta_1}^{A}dq\int_{-\pi}^{\pi}d\theta e^{i\lambda
q\sin\theta}\nu_3(q)h_2(q;\theta).
$$

\bigskip
\noindent\textbf{Step iv)} Consider the first of these integrals
in the limit $A\rightarrow\infty$,

\begin{equation}
-\frac{1}{\lambda^2}
\int_{\beta_1}^{\infty}dq\int_{-\pi}^{\pi}d\theta e^{i\lambda
q\sin\theta}[e^{-i\lambda\tau e^{-q}}-1]\nu_3(q)h_2(q;\theta).
\label{28bis}
\end{equation}
This integral converges because $(e^{-i\lambda\tau e^{-q}}-1)$
falls off exponentially as $q\rightarrow\infty$. Let us define now
the functions $f(q,\theta)=q^{\sigma}(e^{-i\lambda\tau e^{-q}}-1)$
and $g(q,\theta)=q^{-\sigma}e^{i\lambda
q\sin\theta}\nu_3(q)h_2(q;\theta)$ with $\sigma$ chosen so that
$f,g\in L^2\big([\beta_1,\infty)\times[-\pi,\pi)\big)$. For
example we set $\sigma=3$. The H\"{o}lder inequality gives
\begin{eqnarray*}
\bigg| \frac{1}{\lambda^2}
\int_{\beta_1}^{\infty}dq\int_{-\pi}^{\pi}d\theta e^{i\lambda
q\sin\theta}[e^{-i\lambda\tau e^{-q}}-1]\nu_3(q)h_2(q;\theta)
\bigg|\leq
\frac{1}{\lambda^2}\int_{\beta_1}^{\infty}dq\int_{-\pi}^{\pi}d\theta
|f(q,\theta)g(q,\theta)|&&\\
\leq \frac{1}{\lambda^2}
\bigg\{\int_{\beta_1}^{\infty}dq\int_{-\pi}^{\pi}d\theta
|f(q,\theta)|^2
\bigg\}^{\frac{1}{2}}\bigg\{\int_{\beta_1}^{\infty}dq\int_{-\pi}^{\pi}d\theta
|g(q,\theta)|^2 \bigg\}^{\frac{1}{2}}.\hspace{3.5cm}&&
\end{eqnarray*}
The integral involving $|g(q,\theta)|^2$ is convergent and
independent of $\lambda$ and the one involving $|f(q,\theta)|^2$
satisfies
$$
\bigg\{\int_{\beta_1}^{\infty}dq\int_{-\pi}^{\pi}d\theta
|f(q,\theta)|^2 \bigg\}^{\frac{1}{2}}\leq
2\sqrt{2\pi}\bigg\{\int_0^{\infty}dq q^6
\sin^2\left(\frac{\tau\lambda}{2}
e^{-q}\right)\bigg\}^{\frac{1}{2}}=2\sqrt{2\pi}\bigg\{\int_0^1dt
\frac{1}{t}(\log t)^6 \sin^2(\Lambda t)\bigg\}^{\frac{1}{2}},
$$
after changing variables ($t=e^{-q}$) and defining
$\Lambda=\tau\lambda/2$. As we show in Appendix I this last
integral is $O(\log^7\lambda)$ and, hence, the term (\ref{28bis})
is $O(\log^{7/2}\lambda/\lambda^2)$. So in expression
(\ref{improper4}) we obtain
\begin{eqnarray}
\Im\bigg\{\frac{\lambda e^{i\tau\lambda}}{2\pi
R_1}\lim_{A\rightarrow\infty}\bigg[-\frac{i}{\lambda}\int_{-\pi}^{\pi}d\theta
\frac{\sin\theta}{A^2\cos^2\theta+\sin^2\theta}e^{i\lambda
A\sin\theta}\hspace{2.9cm}&&\nonumber\\
+\frac{1}{\lambda^2} \int_{-\pi}^{\pi}d\theta
\frac{A(A^2-4)\sin^2\theta\cos^2\theta-A\sin^4\theta}
{[A^2\cos^2\theta+\sin^2\theta]^3}e^{i\lambda
A\sin\theta}\hspace{.5cm}
&&\label{29bis}\\
-\frac{1}{\lambda^2} \int_{\beta_1}^{A}dq\int_{-\pi}^{\pi}d\theta
e^{i\lambda
q\sin\theta}\nu_3(q)h_2(q;\theta)\bigg]\bigg\}+O\bigg[\frac{1}{\lambda}\bigg]+
O\bigg[\frac{\log^{7/2}\lambda}{\lambda}\bigg].
\hspace{-1.7cm}&&\nonumber
\end{eqnarray}
The reason why we had to integrate by parts in the first step is
to get an additional factor of $\lambda$ dividing the powers of
$\log\lambda$.

\bigskip
\noindent\textbf{Step v)} Using integration by parts twice one can
check that the terms in the square brackets in (\ref{29bis}) can
be written when $A\rightarrow\infty$ as
\begin{equation}
2\pi\lim_{A\rightarrow\infty}\int_{\beta_1}^A dq\,
\nu_3(q)J_0(\lambda q)=\lim_{A\rightarrow\infty}\int_{\beta_1}^A
dq\int_{-\pi}^{\pi}d\theta \,\nu_3(q)e^{i\lambda
q\sin\theta},\label{intJ}
\end{equation}
plus an extra contribution coming from derivatives of $\nu_3(q)$
that does not contribute at this asymptotic order. In Appendix II
we show that (\ref{intJ}) is $O(1/\lambda^R)$ for all $R>0$. We
therefore conclude that (\ref{improper2}) is
$O(\log^{7/2}\lambda/\lambda)$. In our proof we have used
integration by parts twice to arrive at (\ref{29bis}). Actually by
using it repeatedly and applying the five-steps procedure
explained above it is possible to argue that
$\lim_{A\rightarrow\infty}I_3$ is, in fact, $O(1/\lambda^R)$ for
all $R>0$.

We hence conclude that the first terms in the asymptotic expansion
of (\ref{bintegral}) are given by the contributions (\ref{ff1})
and (\ref{ff2}), and therefore by (\ref{lambdasimpt1}) when
$\tau>1$. In order to get the first non-zero term in the
asymptotics for $\tau<1$ it is necessary to perform integration by
parts twice and follow the steps detailed above. This leads to the
contribution for $\tau<1$ shown in (\ref{lambdasimpt2}),
$$
-\frac{1}{2\pi
R_1}\Im\bigg\{\frac{i}{\lambda}\oint_{\gamma}dt\frac{8i\tau
t^2}{(t^2+2i\tau
t-1)^3}\bigg\}=\frac{1}{R_1}\frac{\tau(1+2\tau^2)}{2\lambda(1-\tau^2)^{5/2}}.
$$

The $\lambda\rightarrow0^+$ limit, on the other hand, can be
obtained by using Taylor's theorem to write
$$
e^{-i\tau\lambda e^{-t/\lambda}}=\sum_{k=0}^N
\frac{(-i\tau\lambda)^k}{k!}e^{-\frac{kt}{\lambda}}+
\frac{(-i\tau\lambda)^{(N+1)}}{(N+1)!}e^{-i\xi(t)}e^{-\frac{N+1}{\lambda}t},
$$
with $0<\xi(t)<\tau\lambda e^{-t/\lambda}$. Substituting this in
(\ref{r-int-inf}) (with $\rho=0$) we get
$$
\frac{1}{R_1}\Im\bigg\{e^{i\tau\lambda}\int_0^{\infty}dt
J_0(t)\bigg[\sum_{k=0}^N
\frac{(-i\tau\lambda)^k}{k!}e^{-\frac{kt}{\lambda}}+
\frac{(-i\tau\lambda)^{(N+1)}}{(N+1)!}e^{-i\xi(t)}e^{-\frac{N+1}{\lambda}t}\bigg]
\bigg\}.
$$
The contribution of the last term can be easily bounded
$$
\bigg|\frac{e^{i\tau\lambda}(-i\tau\lambda)^{N+1}}{(N+1)!}
\int_{0}^{\infty}\!\!dt\,J_0(t)e^{-i\xi(t)}e^{-\frac{N+1}{\lambda}t}\bigg|\leq\\
\frac{(\tau\lambda)^{N+1}}{(N+1)!}
\int_0^{\infty}\!\!dt\,e^{-\frac{N+1}{\lambda}t}=
\frac{\tau^{N+1}\lambda^{N+2}}{(N+1)(N+1)!}=O(\lambda^{N+2}).
$$
So we find that the asymptotics of (\ref{r-int-inf}) in this limit
is given by
$$
\frac{\lambda}{R_1}\Im\bigg\{e^{i\tau\lambda}
\sum_{k=0}^{N}\frac{(-i\tau\lambda)^k}{k!\sqrt{k^2+\lambda^2}}\bigg\}+O(\lambda^{N+2}).
$$
\subsubsection{\label{rholambneq0} $\rho\neq0$}

We study now the integral (\ref{trintegral}) for $\rho\neq0$. We
will basically follow the same steps of the $\rho=0$ case so we
will skip some details. We choose $\eta_q=-1$, $\eta_{t_1}=t_1^2$,
and $\eta_{t_2}=t_2^2$, obtaining
$$
\|\nabla\Phi\|^2=\frac{1}{4}\left\{q^2\left[\left( t_1 +
\frac{1}{t_1}\right)^2+\rho^2\left(t_2 + \frac{1}{t_2}\right)^2
\right]-\left[ t_1-\frac{1}{t_1} + \rho\big(t_2-\frac{1}{t_2}\big)
+2i\tau e^{-q} \right]^2\right\}.
$$
As we did above we introduce neutralizers $\nu_j$, $j=1,\,2,\,3$
and write (\ref{trintegral}) as a sum of the tree integrals
\begin{eqnarray*}
&& I_j\equiv-\frac{1}{R_1}\Im\Big\{\frac{\lambda
e^{i\tau\lambda}}{4\pi^2}\int_0^{\infty}dq\oint_{\gamma_1}
dt_1\oint_{\gamma_2}dt_2\frac{\nu_j(q)}{t_1t_2}
e^{\lambda\big[\frac{\scriptstyle
q}{\scriptstyle2}(t_1-\frac{\scriptstyle1}{\scriptstyle
t_1})+\frac{\scriptstyle \rho
q}{\scriptstyle2}(t_2-\frac{\scriptstyle1}{\scriptstyle
t_2})-i\tau e^{-q}\big]}\Big\}.\label{trintegralIj}
\end{eqnarray*}
Starting with $I_1$ we fix $\gamma_{1,2}$ as in fig.
\ref{fig7:gamma}. Integrating by parts we find that only the
boundary term contributes at leading order. This contribution can
be written as
\begin{equation}
\frac{1}{R_1}\Im\bigg\{\frac{1}{2\pi^2}
\oint_{\gamma_1}dt_1\oint_{\gamma_2}dt_2\frac{1}{\rho
t_1(t_2^2-1)+t_2(t_1^2+2i\tau t_1-1)}\bigg\}.\label{freefront}
\end{equation}
This integral can be computed exactly (see Appendix III) and, in
fact, it is equal to the free commutator given in
\cite{BarberoG.:2003ye}; in particular it is zero in the regions
IA and IB of fig. \ref{fig1:regions}. This means that we will have
to use integration by parts again, as in the $\rho=0$ case, to get
the first significant asymptotic term when
$\lambda\rightarrow\infty$ for these regions. After doing this we
get the double integral
\begin{equation}
-\frac{1}{R_1}\Im\bigg\{\frac{2i\tau}{\pi^2\lambda}
\oint_{\gamma_1}dt_1\oint_{\gamma_2}dt_2\frac{t_1^2t_2^2}{[\rho
t_1(t_2^2-1)+t_2(t_1^2+2i\tau t_1-1)]^3}\bigg\}.\label{doble}
\end{equation}
It can be easily checked that this reproduces the result obtained
above for $\rho=0$. This integral, in the considered regions
outside the light cone, can be written in terms of complete
elliptic integrals of the first and second kind, as shown in
Appendix IV. The result is just the $1/\lambda$ contribution
appearing in equation (\ref{IAIB}).

Though it is also possible to compute the integral $I_1$ in the
remaining regions, we will not do so because their contributions
are subdominant with respect to those coming from the critical
points of $I_2$.

Consider then $I_2$ and take as before the unit circumference as
the integration path $\gamma_{1,2}$. We parameterize this curve
according to $t_1(\theta_1)=e^{i\theta_1}$,
$t_2(\theta_2)=e^{i\theta_2}$, $\theta_{1,2}\in(-\pi,\pi]$. Now we
have
$$
i\lambda\phi(q;\theta_1,\theta_2)\equiv\lambda\Phi[q;t_1(\theta_1),t_2(\theta_2)]=
i\lambda[q(\sin\theta_1+\rho\sin\theta_2)-\tau e^{-q}],
$$
so that the critical points are given by the solutions to the
equations
\begin{eqnarray*}
\sin\theta_1+\rho\sin\theta_2+\tau e^{-q}=0,&&\\
q\cos\theta_1=0,&&\\
q\rho\cos\theta_2=0.&&\\
\end{eqnarray*}
Since we are computing $I_2$ we are only interested in solutions
with $q\neq0$. Therefore we must have $\cos\theta_1=0$ and
$\cos\theta_2=0$, i.e. $\theta_1=\pm\frac{\pi}{2}$ and
$\theta_2=\pm\frac{\pi}{2}$. For
$\theta_1=\theta_2=\frac{\pi}{2}$, the remaining equation implies
$1+\rho+\tau e^{-q}=0$, which has no solutions for
$q\in\mathbb{R}$. If $\theta_1=-\theta_2=\frac{\pi}{2}$ we get
$1-\rho+\tau e^{-q}=0$; this equation has solutions
$q=\log\frac{\tau}{\rho-1}$ in the integration region $q>0$ only
if $\tau>\rho-1>0$. If $\theta_1=-\theta_2=-\frac{\pi}{2}$, we
must have $-1+\rho+\tau e^{-q}=0$, and there exists a critical
point $q=\log\frac{\tau}{1-\rho}$ in the integration region only
if $\tau>1-\rho>0$. Finally if $\theta_1=\theta_2=-\frac{\pi}{2}$
we obtain $-1-\rho+\tau e^{-q}=0$, which is solved by
$q=\log\frac{\tau}{1+\rho}$; in this case $q>0$ only when
$\tau>1+\rho$. As we see when $(\tau,\rho)$ are in the regions IA
and IB of fig. \ref{fig1:regions} there are no critical points for
$I_2$, if $(\tau,\rho)$ is in region II there is only one critical
point, and if $(\tau,\rho)$ is in region III there are two
critical points. The contribution of these critical points to the
asymptotics of $I_2$, that can be obtained with the standard
formulas \cite{Handels}, are the following in regions II and III
respectively:
$$
\frac{1}{R_1} \Im\bigg\{\frac{e^{-i\frac{\pi}{4}}
e^{i\lambda[\tau+|\rho-1|(1+\log\frac{\tau}{|\rho-1|})]}}
{\sqrt{2\pi\lambda \rho|1-\rho|}\log\frac{\tau}{|1-\rho|}}\bigg\}+
O\bigg[\frac{1}{\lambda^{3/2}}\bigg],
$$
$$
\frac{1}{R_1} \Im\bigg\{\frac{e^{-i\frac{\pi}{4}}
e^{i\lambda[\tau-|\rho-1|(1+\log\frac{\tau}{|\rho-1|})]}}
{\sqrt{2\pi\lambda \rho|1-\rho|}\log\frac{\tau}{|1-\rho|}}+
\frac{e^{i\frac{\pi}{4}}e^{i\lambda[\tau+(\rho+1)(\log\frac{1+\rho}{\tau}-1)]}}
{\sqrt{2\pi\lambda \rho(1+\rho)}\log\frac{\tau}{1+\rho}}\bigg\}+
O\bigg[\frac{1}{\lambda^{3/2}}\bigg].
$$
These, together with the contribution of the boundary term,
provide (\ref{II}) and (\ref{III}).

Finally it is possible to prove that
$\lim_{A\rightarrow\infty}I_3$ gives no contribution by
essentially following the same steps as when $\rho=0$.

It is interesting to comment on the singularities that appear in
the borders between the different regions I, II, III, and the axis
$\rho=0$. Their presence is a manifestation of the fact that the
asymptotic behavior in $\lambda$ changes abruptly between adjacent
regions. Notice, nonetheless, that the behavior is continuous in
the border between the portion of the axis with $\tau<1$ and
region IA, where the leading terms of the asymptotic expansion in
$\lambda$ are both $O(1/\lambda)$.

To conclude let us point out that the limit
$\lambda\rightarrow0^+$ can be obtained as in the $\rho=0$ case.
The result is
$$
\frac{1}{\pi R_1\sqrt{\rho}}\Im\bigg\{e^{i\tau\lambda}
\sum_{k=0}^{N}\frac{(-i\tau\lambda)^k}{k!}
Q_{-\frac{1}{2}}\bigg[\frac{\lambda^2(1+\rho^2)+k^2}
{2\rho\lambda^2}\bigg]\bigg\}+O(\lambda^{N+2}).
$$


\section{\label{Conclusions}Conclusions and comments}

We have analyzed in detail the issue of microcausality for
linearly polarized cylindrical waves by looking at the commutator
of the axially symmetric scalar field that encodes the physical
degrees of freedom of the system. We have been able to show
several interesting effects that appear in the model. The first is
a smearing of the cylindrical light cones. This is especially
obvious if one studies the behavior of the commutator (divided by
$8iG$) at two points with coordinates $(t_1,R_1)$ and $(t_2,R_2)$
when $\lambda$ (the quotient of $R_1$ and the Planck length) is
large. If $R_2$ is not too close to the axis one gets, in this
limit, the discontinuous, cylindrical light cone structure defined
by the free commutator. In particular, outside the region defined
by this light cone the commutator is zero and, hence, observables
would commute as in ordinary perturbative QFT. If, instead, one
looks at the behavior when $R_2=0$ one finds a peculiar behavior
as $\lambda\rightarrow\infty$; now superimposed to the free
contribution, the commutator shows a characteristic oscillating
behavior when the variable $\tau=(t_2-t_1)/4G$ grows. The
frequency of this oscillation is controlled by the value of
$\lambda$ but the amplitude is \emph{independent} of it, and turns
out to decrease very slowly with $\tau$ in such a way that one
recovers the value given by the free commutator only for very
large values of $\tau$. Nonetheless, there is a sense in which the
free propagator is recovered if one averages over time intervals
sufficiently larger than $4G$.

In our study we have had to find the most efficient methods to
obtain the relevant asymptotic behaviors. Though this has not
required the introduction of completely novel techniques it has
been necessary to extend to our situation the usual Mellin
transform methods and those employed for the analysis of multiple
integrals. We have needed also to take into account that the
integrals that define the commutator are, actually, improper.

Several issues are worth discussing in more detail. One is the
problem of introducing regulators in the field operators and
compare the results with those derived here. Though it is clear
that the approximation provided by the improper integrals
considered in this work must be good in certain regimes, at some
point the existence of a physical cut-off might manifest itself in
the behavior of the field commutator, especially in the asymptotic
regimes that we have explored. Another task that can be confronted
with the techniques that we have developed here is the computation
of other matrix elements for the commutator. It would also be
interesting to consider other Green functions and some related
objects, such as S matrix elements. We plan to do that in the
future.


\section{\label{A1}Appendix I}

We want to investigate the asymptotic behavior of
$\int_{0}^{1}dt\frac{(\log t)^m}{t}\sin^2\Lambda t$,
$m\in\mathbb{N}$, when $\Lambda\rightarrow\infty$. This can be
found by a straightforward use of the Mellin-Parseval formula
(\ref{th3}) in order to get a Mellin-Barnes representation for the
integral. Defining $h(t)=\sin^2t$ and $f(t)=\frac{(\log t)^m}{t}$
we get the Mellin transforms
\begin{eqnarray*}
&&M[h;z]=-\frac{\Gamma(z)\cos\frac{\pi
z}{2}}{2^{1+z}}\quad-2<\Re(z)<0, \\
&&M[f;1-z]=-\frac{\Gamma(m+1)}{z^{m+1}}\quad\quad\quad1<\Re(z),
\end{eqnarray*}
so that
$$
\int_{0}^{1}dt\frac{(\log t)^m}{t}\sin^2\Lambda t=\frac{m!}{2\pi
i}\int_{c-i\infty}^{c+i\infty}\!\!\!dz\,\Lambda^{-z}\frac{\Gamma(z)\cos\frac{\pi
z}{2}}{2^{1+z}z^{m+1}}\quad\quad c\in(-2,0).
$$
The integrand in this last expression can be analytically extended
as a meromorphic function to $\mathbb{C}$ with poles in $z=-2n$,
$n=0,\,1,\ldots$. By displacing the integration contour to the
right of the pole at $z=0$, we get
\begin{equation}
\int_{0}^{1}dt\frac{(\log t)^m}{t}\sin^2\Lambda
t=-{\textrm{res}}\bigg[\Lambda^{-z}\frac{m!\Gamma(z)\cos\frac{\pi
z}{2}}{2^{1+z}z^{m+1}},z=0\bigg]+\frac{m!}{2\pi
i}\int_{x_0-i\infty}^{x_0+i\infty}\!\!\!dz\,\Lambda^{-z}\frac{\Gamma(z)\cos\frac{\pi
z}{2}}{2^{1+z}z^{m+1}}\label{ap1}
\end{equation}
($x_0>0$) whenever the integral converges and provided we can
neglect the contributions from the segments needed to close the
integration contour at large values of $\Im(z)$. Writing $z=x_0+i
y$ we have
$$
\left|\frac{m!}{2\pi
i}\int_{x_0-i\infty}^{x_0+i\infty}\!\!\!dz\,\Lambda^{-z}\frac{\Gamma(z)\cos\frac{\pi
z}{2}}{2^{1+z}z^{m+1}}\right|\leq\frac{m!}{4\pi(2\Lambda)^{x_0}}\int_{-\infty}^{\infty}\!\!dy\,\left|
\frac{\Gamma(x_0+iy)\cos\frac{\pi}{2}(x_0+iy)}{(x_0+iy)^{m+1}}
\right|,
$$
and using
$\lim_{|y|\rightarrow\infty}\frac{1}{\sqrt{2\pi}}|\Gamma(x+iy)|e^{\pi|y|/2}|y|^{-x+1/2}=1$
we find that
$$
\left|
\frac{\Gamma(x_0+iy)\cos\frac{\pi}{2}(x_0+iy)}{(x_0+iy)^{m+1}}
\right|\sim \sqrt{\frac{\pi}{2}}|y|^{x_0-m-3/2}.
$$
Hence, the integral in (\ref{ap1}) is absolutely convergent if
$x_0<m+\frac{1}{2}$ and its contribution is
$O\big[\frac{1}{\Lambda^{x_0}}\big]$. It is straightforward to
show that the residue at the pole $z=0$ is an $m+1$ degree
polynomial in $\log\Lambda$ with the higher degree term given by
$\frac{(-1)^{m+1}}{2(m+1)}(\log\Lambda)^{m+1}$. This proves that,
for $m\in\mathbb{N}$,
$$\int_{0}^{1}dt\frac{(\log t)^m}{t}\sin^2\Lambda t=O\big[(\log\Lambda)^{(m+1)}\big].$$


\section{\label{A2}Appendix II}

Let us analyze the asymptotic behavior of
$2\pi\int_{\beta_1}^{\infty}dq\,\nu_3(q)J_0({\lambda q})$ when
$\lambda\rightarrow\infty$. After changing variables according to
$k=\lambda q$ this integral becomes
$$
\frac{2\pi}{\lambda}\int_{\lambda\beta_1}^{\infty}dk\,\nu_3\left(\frac{k}{\lambda}\right)J_0(k).
$$
Taking into account that
$J_0(k)=\bigg[-\frac{1}{k}\frac{d\,\,}{dk}-\frac{d^2}{dk^2}\bigg]J_0(k)$,
this can be written as
$$
\frac{2\pi}{\lambda}\int_{\lambda\beta_1}^{\infty}dk\,\nu_3\left(\frac{k}{\lambda}\right)
\bigg[-\frac{1}{k}\frac{d\,\,}{dk}-\frac{d^2}{dk^2}\bigg]^NJ_0(k)
$$
with $N\in\mathbb{N}$. It is straightforward to prove by induction
that
$$\bigg[-\frac{1}{k}\frac{d\,\,}{dk}-\frac{d^2}{dk^2}\bigg]^N=
\sum_{j=1}^{2N}\frac{a_j}{k^{2N-j}} \frac{d^j}{dk^j}$$ for certain
coefficients $a_j\in\mathbb{Z}$. Since all the derivatives of
$\frac{1}{k^{2N-j}}\nu_3\big(\frac{k}{\lambda}\big)$ at
$k=\lambda\beta_1$ cancel, by integrating by parts $j$ times and
changing variables according to $k=q\lambda$ we obtain for our
integral the following expression
$$
\frac{2\pi}{\lambda}\sum_{j=1}^{2N}(-1)^ja_j\int_{\lambda\beta_1}^{\infty}dk\,
\frac{d^j}{dk^j}\bigg[\frac{1}{k^{2N-j}}\nu_3\left(\frac{k}{\lambda}\right)\bigg]
J_0(k)=\frac{2\pi}{\lambda^{2N}}\sum_{j=1}^{2N}(-1)^ja_j\int_{\beta_1}^{\infty}dq\,
\frac{d^j}{dq^j}\bigg[\frac{\nu_3(q)}{q^{2N-j}}\bigg] J_0(\lambda
q).
$$
Using that $J_0(x)\leq 1$ for all $x\in\mathbb{R}$, we hence get
$$
\left|\frac{2\pi}{\lambda}\int_{\lambda\beta_1}^{\infty}
dk\,\nu_3\left(\frac{k}{\lambda}\right)J_0(k)\right|\leq
\frac{2\pi}{\lambda^{2N}}\sum_{j=1}^{2N}|a_j|\int_{\beta_1}^{\infty}dq\,
\left|\frac{d^j}{dq^j}\bigg[\frac{\nu_3(q)}{q^{2N-j}}\bigg]\right|,
$$
with all the integrals in the last sum being convergent. We
conclude that, for all $N\in\mathbb{N}$,
$$
\frac{2\pi}{\lambda}\int_{\lambda\beta_1}^{\infty}dk\,
\nu_3\left(\frac{k}{\lambda}\right)J_0(k)=O\bigg[\frac{1}{\lambda^{2N}}\bigg].
$$


\section{\label{A3}Appendix III}

\begin{figure}
\hspace{0cm}\includegraphics[width=10.5cm]{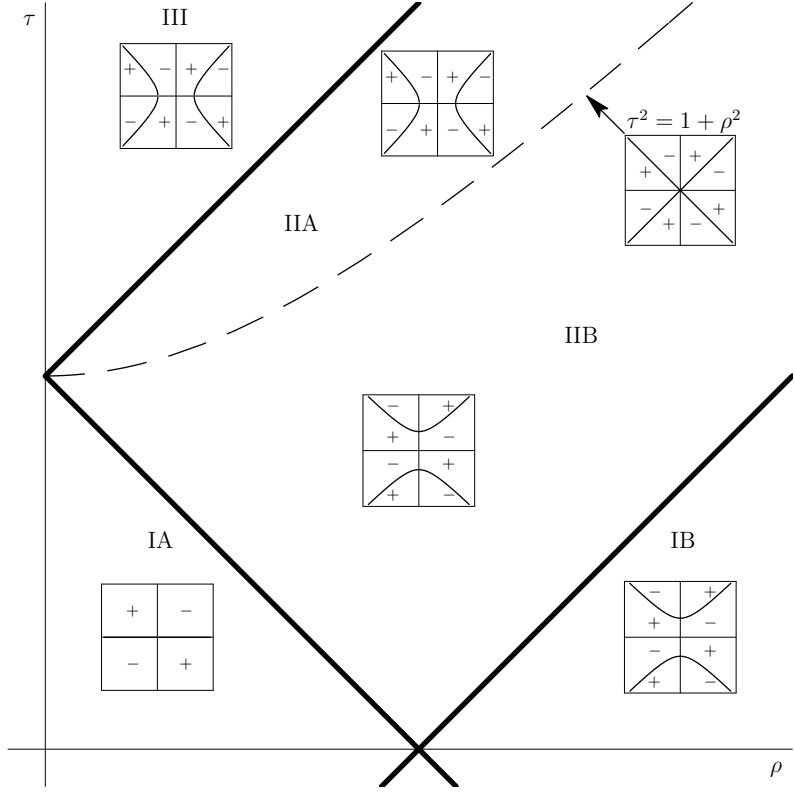}
\caption{Sign of $\Re[\frac{\sqrt{Q_1(s)Q_2(s)}}{i(s^2-1)}]$ for
every choice of $\rho$ and $\tau$. Each inset shows this sign for
$s=x+iy\in\mathbb{C}$. The hyperbolas that separate some of the
regions in these insets have the equation
$x^2-y^2=\frac{1+\rho^2-\tau^2}{1-(\rho+\tau)^2}$.}
\label{fig8:sign}
\end{figure}

In this appendix we compute the integral (\ref{freefront}). It is
possible to show that choosing $\gamma_{1,2}$ as in fig.
\ref{fig7:gamma} its denominator is always different from zero
except for those exceptional values of $\rho$ and $\tau$
corresponding to the borders between regions I, II, and III. We
can use Fubini's theorem and exchange orders of integration. To
integrate in $t_1$ we fix a value of $t_2$ on $\gamma$. The poles
of the integrand are
$$
t_1^{\pm}=\frac{1}{2t_2}\bigg[\rho-\rho t_2^2-2i\tau
t_2\pm\sqrt{4t^2_2+(-\rho+\rho t_2^2+2i\tau t_2)^2}\bigg].
$$
One of them is always inside $\gamma$ and the other outside. For a
fixed value of $t_2$ we find that $t_1^-$ is inside $\gamma$ when
$$
\Re\bigg[\frac{1}{t_2}\sqrt{4t^2_2+(-\rho+\rho t_2^2+2i\tau
t_2)^2}\bigg]>0
$$
or in certain exceptional cases when $\Re(t_1^-)=0$. Analogously
$t_1^+$ is inside $\gamma$ when
$$
\Re\bigg[\frac{1}{t_2}\sqrt{4t^2_2+(-\rho+\rho t_2^2+2i\tau
t_2)^2}\bigg]<0
$$
or, again, in certain exceptional occasions with $\Re(t_1^+)=0$
(in this last case $t_1^-$ is outside $\gamma$). The residues at
these poles are
\begin{eqnarray*}
t_1^+\longrightarrow\frac{1}{\sqrt{4t^2_2+(-\rho+\rho t_2^2+2i\tau
t_2)^2}},&&\\
t_1^-\longrightarrow\frac{-1}{\sqrt{4t^2_2+(-\rho+\rho
t_2^2+2i\tau t_2)^2}},&&
\end{eqnarray*}
and integrating in $t_1$, (\ref{freefront}) becomes
\begin{equation}
\Im\bigg\{-\frac{i}{\pi R_1}\oint_{\gamma}dt\,\frac{
\textrm{sgn}\,\Re[\frac{1}{t}\sqrt{4t^2+(-\rho+\rho t^2+2i\tau
t)^2}]}{\sqrt{4t^2+(-\rho+\rho t^2+2i\tau t)^2}}
\bigg\}.\label{freeintegral1}
\end{equation}
Let us define $A=-1+\rho-\tau$, $B=1+\rho+\tau$, $C=1+\rho-\tau$,
and $D=-1+\rho+\tau$. It is not difficult to check that
$$
4t^2+(-\rho+\rho t^2+2i\tau
t)^2=\frac{1}{4}[A(t-i)^2+B(t+i)^2][C(t-i)^2+D(t+i)^2],
$$
and, hence, the integral (\ref{freeintegral1}) is
\begin{equation}
\Im\bigg\{-\frac{2i}{\pi
R_1}\oint_{\gamma}dt\frac{\textrm{sgn}\,\Re[\frac{1}{t}
\sqrt{[A(t-i)^2+B(t+i)^2][C(t-i)^2+D(t+i)^2]}}
{\sqrt{[A(t-i)^2+B(t+i)^2][C(t-i)^2+D(t+i)^2]}}
\bigg\}.\label{freeintegral2}
\end{equation}
In the following we restrict $\gamma$ to be a positively oriented
circumference going through $\pm i$ and parameterize it as
$z(\theta)=-a+\sqrt{1+a^2}e^{i\theta}$, with $\theta\in[0,2\pi)$
and $a>0$ a constant. Changing variables according to the
M\"{o}bius transformation $s(t)=\frac{t+i}{t-i}$ and denoting
$Q_1(s)=A+Bs^2$, $Q_2(s)=C+Ds^2$ we can write
(\ref{freeintegral2}) in the form
\begin{equation}
\Im\bigg\{-\frac{1}{\pi
R_1}\int_{s(\gamma)}\!\!\!ds\,\frac{\textrm{sgn}\,
\Re\left[\frac{s-1}{i(s+1)}\sqrt{\frac{Q_1(s)Q_2(s)}{(s-1)^4}}\right]}
{(s-1)^2\sqrt{\frac{Q_1(s)Q_2(s)}{(s-1)^4}}}\bigg\}=
\Im\bigg\{-\frac{1}{\pi
R_1}\int_{s(\gamma)}\!\!\!ds\,\frac{\textrm{sgn}\,
\Re[\frac{\sqrt{Q_1(s)Q_2(s)}}{i(s^2-1)}]}{\sqrt{Q_1(s)Q_2(s)}}\bigg\}\equiv
\Im(I)\label{freeintegral3}
\end{equation}
where $s(\gamma)$ --the image of the circumference $\gamma$-- is a
straight line through the origin with slope $-\frac{1}{a}<0$. The
sign in the numerator in the last integral, for each
$s\in\mathbb{C}$, is shown in fig. \ref{fig8:sign}  in the
different regions in the $(\rho,\tau)$ parameter space.
We now proceed to compute the integral $I$ appearing in
(\ref{freeintegral3}) for the regions IA, IB, IIA, IIB, and III,
and for the hyperbola $\tau^2=\rho^2+1$.

\bigskip
\noindent\textbf{Region IA:} Here $A<0$, $B>0$, $C>0$, $D<0$, and
the sign in the integrand is positive. We choose $a\rightarrow0$
and parameterize $s(\gamma)$ as $s=\sigma(i-\epsilon)$
[$\sigma\in\mathbb{R}$, $\epsilon\rightarrow 0^+$]. We have
$$
I=-\frac{i}{\pi
R_1}\int_{\mathbb{R}}\frac{d\sigma}{\sqrt{(A-B\sigma^2)(C-D\sigma^2)-i\varepsilon}}
$$
with $\varepsilon\rightarrow0^+$. Recalling our choice of branch
for the square root, changing variables according to
$u=\sqrt{-\frac{B}{A}}\sigma$, and noticing that
$0<\frac{AD}{BC}<1$, this integral can be written as
\begin{equation}
I=\frac{1}{\pi
R_1}\frac{1}{\sqrt{BC}}\int_{\mathbb{R}}\frac{du}{\sqrt{(1+u^2)(1+\frac{AD}{BC}u^2)}}=
\frac{2}{\pi R_1}\frac{1}{\sqrt{(1+\rho)^2-\tau^2}}
K\bigg(\sqrt{\frac{4\rho}{(1+\rho)^2-\tau^2}}\bigg).\label{36bis}
\end{equation}

\bigskip
\noindent\textbf{Region IB:} Here $A>0$, $B>0$, $C>0$, $D>0$, and
the sign in the integrand is positive. We choose
$a\rightarrow\infty$ and parameterize $s(\gamma)$ as
$s=\sigma(-1+i\epsilon)$ [$\sigma\in\mathbb{R}$,
$\epsilon\rightarrow 0^+$]. Then
$$
I=\frac{i}{\pi
R_1}\int_{\mathbb{R}}\frac{d\sigma}{\sqrt{(A+B\sigma^2)(C+D\sigma^2)-i\varepsilon}}
$$
with $\varepsilon\rightarrow0^+$. After the change of variables
$u=\sqrt{\frac{B}{A}}\sigma$ (and since $0<\frac{AD}{BC}<1$) this
integral becomes exactly (\ref{36bis}).

\bigskip
\noindent\textbf{Region IIA:} Here $A<0$, $B>0$, $C>0$, $D>0$, and
the sign in the integrand is negative. Letting $a\rightarrow0$ and
parameterizing $s(\gamma)$ as $s=\sigma(i-\epsilon)$
[$\sigma\in\mathbb{R}$, $\epsilon\rightarrow 0^+$] we obtain (with
$\varepsilon\rightarrow0^+$)
$$
I=\frac{i}{\pi
R_1}\int_{\mathbb{R}}\frac{d\sigma}{\sqrt{(A-B\sigma^2)(C-D\sigma^2)+i\varepsilon}}.
$$
This last integral can be split in two pieces:
\begin{eqnarray*}
&&I=\frac{2}{\pi
R_1}\int_0^{\sqrt{\frac{C}{D}}}\frac{d\sigma}{\sqrt{(-A+B\sigma^2)(C-D\sigma^2)}}+
\frac{2i}{\pi
R_1}\int_{\sqrt{\frac{C}{D}}}^{\infty}\frac{d\sigma}{\sqrt{(A-B\sigma^2)(C-D\sigma^2)}}.
\end{eqnarray*}
We find then (see formulas 3.152-3 and 3.152-6 of
\cite{Gradshteyn})
\begin{equation}
I=\frac{1}{\pi
R_1\sqrt{\rho}}K\bigg(\sqrt{\frac{(1+\rho)^2-\tau^2}{4\rho}}\bigg)+\frac{i}{\pi
R_1\sqrt{\rho}}K\bigg(\sqrt{\frac{\tau^2-(\rho-1)^2}{4\rho}}\bigg).\label{regII}
\end{equation}

\bigskip
\noindent\textbf{Region IIB:} Here $A<0$, $B>0$, $C>0$, $D>0$, and
the sign in the integrand is positive. With the choice
$a\rightarrow\infty$ and the parametrization
$s=\sigma(-1+i\epsilon)$ [$\sigma\in\mathbb{R}$] for $s(\gamma)$
we arrive at
$$
I=\frac{1}{\pi
R_1}\int_{\mathbb{R}}\frac{d\sigma}{\sqrt{(A+B\sigma^2)(C+D\sigma^2)-i\varepsilon}}
$$
with $\varepsilon\rightarrow0^+$. This can be written as
\begin{eqnarray*}
&&I=\frac{2i}{\pi
R_1}\int_0^{\sqrt{-\frac{A}{B}}}\frac{d\sigma}{\sqrt{-(A+B\sigma^2)(C+D\sigma^2)}}+
\frac{2}{\pi
R_1}\int_{\sqrt{-\frac{A}{B}}}^{\infty}\frac{d\sigma}{\sqrt{(A+B\sigma^2)(C+D\sigma^2)}},
\end{eqnarray*}
which gives again (\ref{regII}).

\bigskip
\noindent\textbf{Boundary between regions IIA and IIB:} This is
the hyperbola $\tau^2=1+\rho^2$. Parameterizing $s(\gamma)$ as
$s=\sigma(i-\epsilon)$ [$\sigma\in\mathbb{R}$,
$\epsilon\rightarrow 0^+$] we get
$$
I=\frac{i}{\pi
R_1}\int_{\mathbb{R}}\frac{d\sigma}{\sqrt{(A-B\sigma^2)(C-D\sigma^2)+i\varepsilon}}
$$
(with $\varepsilon\rightarrow 0^+$) which reduces to the previous
two cases. Substituting $\tau^2=1+\rho^2$ we thus obtain
$$
I=\frac{1+i}{\pi R_1\sqrt{\rho}}K\left(\frac{1}{\sqrt{2}}\right)=
\frac{1+i}{4\pi
R_1\sqrt{\pi\rho}}\bigg[\Gamma\left(\frac{1}{4}\right)\bigg]^2.
$$

\bigskip
\noindent\textbf{Region III:} Here $A<0$, $B>0$, $C<0$, $D>0$, and
the sign in the integrand is negative. Letting $a\rightarrow0$ and
parameterizing $s(\gamma)$ as $s=\sigma(i-\epsilon)$
[$\sigma\in\mathbb{R}$, $\epsilon\rightarrow 0^+$] we obtain (with
$\varepsilon\rightarrow0^+$)
$$
I=\frac{i}{\pi
R_1}\int_{\mathbb{R}}\frac{d\sigma}{\sqrt{(A-B\sigma^2)(C-D\sigma^2)-i\varepsilon}}.
$$
With the change of variables $u=\sqrt{-\frac{D}{C}}\sigma$ and
noticing that $0<\frac{BC}{AD}<1$, this becomes
$$
I=\frac{i}{\pi
R_1}\frac{1}{\sqrt{-AD}}\int_{\mathbb{R}}\frac{du}{\sqrt{(1+u^2)(1+\frac{BC}{AD}u^2)}}=
\frac{2i}{\pi
R_1}\frac{1}{\sqrt{\tau^2-(1-\rho)^2}}K\bigg(\sqrt{\frac{4\rho}{\tau^2-(1-\rho)^2}}\bigg).
$$


\section{\label{A4}Appendix IV}
We want to compute the double integral (\ref{doble}). In order to
do this we first integrate in $t_1$. Fixing $t_2$ we find that the
poles coincide with those of (\ref{freefront}). The arguments
showing whether they are inside or outside $\gamma$ are the same
as in Appendix III. The residues are now
\begin{eqnarray*}
t_1^+\longrightarrow\frac{t_2^2[\rho^2(t_2^2-1)^2+4i\rho\tau
t_2(t_2^2-1)-2(1+2\tau^2)t_2^2]}{[4t^2_2+(-\rho+\rho t_2^2+2i\tau
t_2)^2]\sqrt{4t^2_2+(-\rho+\rho t_2^2+2i\tau t_2)^2}},&&\\
t_1^-\longrightarrow-\frac{t_2^2[\rho^2(t_2^2-1)^2+4i\rho\tau
t_2(t_2^2-1)-2(1+2\tau^2)t_2^2]}{[4t^2_2+(-\rho+\rho t_2^2+2i\tau
t_2)^2]\sqrt{4t^2_2+(-\rho+\rho t_2^2+2i\tau t_2)^2}}.&&
\end{eqnarray*}
Integrating in $t_1$, (\ref{doble}) becomes then
\begin{equation}
\Im\bigg\{-\frac{4\tau}{\pi R_1\lambda}\oint_{\gamma}dt
\frac{\textrm{sgn}\,\Re\big[\frac{1}{t}\sqrt{4t^2+(-\rho+\rho
t^2+2i\tau t)^2}\big][\rho^2(t^2-1)^2+4i\rho\tau
t(t^2-1)-2(1+2\tau^2)t^2]t^2}{[4t^2+(-\rho+\rho t^2+2i\tau
t)^2]^2\sqrt{4t^2+(-\rho+\rho t^2+2i\tau t)^2}}
\bigg\}.\label{doble1}
\end{equation}
The same M\"{o}bius transformation used above allows us to write
(\ref{doble1}) as
$$
\Im\bigg\{-\frac{i\tau}{2\pi
R_1\lambda}\int_{s(\gamma)}\!\!\!ds\,\textrm{sgn}\,
\Re\bigg[\frac{\sqrt{Q_1(s)Q_2(s)}}{i(s^2-1)}\bigg]
\frac{(s^2-1)^2[2\rho^2(1+s^2)^2+4\rho\tau(s^4-1)+(1+2\tau^2)(s^2-1)^2]}
{Q_1^2(s)Q_2^2(s)\sqrt{Q_1(s)Q_2(s)}}\bigg\}.
$$
We compute it only in regions IA and IB, remembering that the sign
in the integrand is that shown in fig. \ref{fig8:sign}.

\bigskip
\noindent\textbf{Region IA:} Here $A<0$, $B>0$, $C>0$, $D<0$, and
the sign is positive. We choose $a\rightarrow0$ and parameterize
$s(\gamma)$ as $s=\sigma(i-\epsilon)$ [$\sigma\in\mathbb{R}$,
$\epsilon\rightarrow 0^+$] getting
$$
\Im\bigg\{\frac{i\tau}{2\pi R_1\lambda}\int_{\mathbb{R}}d\sigma
\frac{(1+\sigma^2)^2[2\rho^2(1-\sigma^2)^2+
4\rho\tau(\sigma^4-1)+(1+2\tau^2)(\sigma^2+1)^2]}
{(A-B\sigma^2)^2(C-D\sigma^2)^2\sqrt{-(A-B\sigma^2)(C-D\sigma^2)}}\bigg\}.
$$
The integrand consists of a rational function of $\sigma$ divided
by the square root of a fourth degree polynomial and the
integration extends over the real axis; hence, it can be written
in terms of complete elliptic integrals. The way to proceed now is
to perform a partial fraction decomposition of the rational part
and write it as a sum of integrals of the four following types
$$
\int_{\mathbb{R}}\frac{dx}{\sqrt{(a^2+x^2)(b^2+x^2)^3}},
\int_{\mathbb{R}}\frac{dx}{\sqrt{(a^2+x^2)^3(b^2+x^2)}},
\int_{\mathbb{R}}\frac{dx}{\sqrt{(a^2+x^2)(b^2+x^2)^5}},
\int_{\mathbb{R}}\frac{dx}{\sqrt{(a^2+x^2)^5(b^2+x^2)}}.
$$
These integrals can be found in \cite{Gradshteyn} (see equations
3.158-2, 3.158-4, 3.162-4, and 3.162-2 in that reference). The
result is the $1/\lambda$ contribution appearing in equation
(\ref{IAIB}).

\bigskip
\noindent\textbf{Region IB:} Here $A>0$, $B>0$, $C>0$, $D>0$, and
the sign in the integrand is positive. With the choice
$a\rightarrow0$ and the parametrization $s=\sigma(-1+i\epsilon)$
[$\sigma\in\mathbb{R}$, $\epsilon\rightarrow 0^+$] we obtain
$$
\Im\bigg\{\frac{i\tau}{2\pi R_1\lambda}\int_{\mathbb{R}}d\sigma
\frac{(\sigma^2-1)^2[2\rho^2(1+\sigma^2)^2+
4\rho\tau(\sigma^4-1)+(1+2\tau^2)(\sigma^2-1)^2]}
{(A+B\sigma^2)^2(C+D\sigma^2)^2\sqrt{(A+B\sigma^2)(C+D\sigma^2)}}\bigg\}.
$$
It can be shown that this integral gives again the result found in
region IA.

\begin{acknowledgments}
This work was supported by the Spanish MCYT under the research
projects BFM2001-0213 and BFM2002-04031-C02-02.
\end{acknowledgments}


\end{document}